\documentclass[twocolumn]{aastex63}

\usepackage{epsfig}
\usepackage{soul}
\usepackage{color}
\usepackage{hyperref}
\usepackage[percent]{overpic}

\usepackage{mathtools}
\usepackage{enumerate}
\usepackage{ulem}

\newcommand{\chandra}{\textit{Chandra}}

\newcommand{\xmm}{\textit{XMM-Newton}}

\newcommand{\nustar}{\textit{NuSTAR}}

\newcommand{\ls}{\ensuremath{L_{\odot}}}

\newcommand{\sna}{SN~1987A}

\shorttitle{Additional evidence for a pulsar wind nebula in SN 1987A}
\shortauthors{Greco et al.}

\graphicspath{{./}{figures/}}

\begin{document}

\title{Additional evidence for a pulsar wind nebula in the heart of SN 1987A \\from multi-epoch X-ray data and MHD modeling}

\correspondingauthor{Emanuele Greco}
\email{e.greco@uva.nl }

\author[0000-0001-5792-0690]{Emanuele Greco}
\affiliation{Anton Pannekoek Institute for Astronomy, University of Amsterdam, Science Park 904, 1098 XH Amsterdam, The Netherlands}
\affiliation{INAF-Osservatorio Astronomico di Palermo, Piazza del Parlamento 1, 90134 Palermo, Italy}
\affiliation{GRAPPA, University of Amsterdam, Science Park 904, 1098 XH Amsterdam, The Netherlands}
\affiliation{Universit\`a degli Studi di Palermo, Dipartimento di Fisica e Chimica E. Segr\`e, Piazza del Parlamento 1, 90134 Palermo, Italy}

\author{Marco Miceli}
\affiliation{Universit\`a degli Studi di Palermo, Dipartimento di Fisica e Chimica E. Segr\`e, Piazza del Parlamento 1, 90134 Palermo, Italy}
\affiliation{INAF-Osservatorio Astronomico di Palermo, Piazza del Parlamento 1, 90134 Palermo, Italy}

\author{Salvatore Orlando}
\affiliation{INAF-Osservatorio Astronomico di Palermo, Piazza del Parlamento 1, 90134 Palermo, Italy}

\author{Barbara Olmi}
\affiliation{INAF-Osservatorio Astronomico di Palermo, Piazza del Parlamento 1, 90134 Palermo, Italy}

\author{Fabrizio Bocchino}
\affiliation{INAF-Osservatorio Astronomico di Palermo, Piazza del Parlamento 1, 90134 Palermo, Italy}

\author{Shigehiro Nagataki}
\affiliation{Astrophysical Big Bang Laboratory (ABBL), RIKEN Cluster for Pioneering Research, 2-1 Hirosawa, Wako, Saitama 351-0198, Japan}
\affiliation{RIKEN Interdisciplinary Theoretical and Mathematical Sciences Program (iTHEMS), 2-1 Hirosawa, Wako, Saitama 351-0198, Japan}

\author{Lei Sun}
\affiliation{Department of Astronomy, Nanjing University, Nanjing 210023, People’s Republic of China}
\affiliation{Anton Pannekoek Institute for Astronomy, University of Amsterdam, Science Park 904, 1098 XH Amsterdam, The Netherlands}

\author{Jacco Vink}
\affiliation{Anton Pannekoek Institute for Astronomy, University of Amsterdam, Science Park 904, 1098 XH Amsterdam, The Netherlands}
\affiliation{GRAPPA, University of Amsterdam, Science Park 904, 1098 XH Amsterdam, The Netherlands}

\author{Vincenzo Sapienza}
\affiliation{Universit\`a degli Studi di Palermo, Dipartimento di Fisica e Chimica E. Segr\`e, Piazza del Parlamento 1, 90134 Palermo, Italy}
\affiliation{INAF-Osservatorio Astronomico di Palermo, Piazza del Parlamento 1, 90134 Palermo, Italy}

\author{Masaomi Ono}
\affiliation{Astrophysical Big Bang Laboratory (ABBL), RIKEN Cluster for Pioneering Research, 2-1 Hirosawa, Wako, Saitama 351-0198, Japan}
\affiliation{RIKEN Interdisciplinary Theoretical and Mathematical Sciences Program (iTHEMS), 2-1 Hirosawa, Wako, Saitama 351-0198, Japan}

\author{Akira Dohi}
\affiliation{Department of Physics, Kyushu University, 744 Motooka, Nishi-Ku, Fukuoka Fukuoka 819-0395, Japan}
\affiliation{RIKEN Interdisciplinary Theoretical and Mathematical Sciences Program (iTHEMS), 2-1 Hirosawa, Wako, Saitama 351-0198, Japan}

\author{Giovanni Peres}
\affiliation{Universit\`a degli Studi di Palermo, Dipartimento di Fisica e Chimica E. Segr\`e, Piazza del Parlamento 1, 90134 Palermo, Italy}
\affiliation{INAF-Osservatorio Astronomico di Palermo, Piazza del Parlamento 1, 90134 Palermo, Italy}

\begin{abstract}
Since the day of its explosion, supernova (SN) 1987A has been closely monitored to study its evolution and to detect its central compact relic. In fact, the formation of a neutron star is strongly supported by the detection of neutrinos from the SN. However, besides the detection in the Atacama Large Millimeter/submillimeter Array (ALMA) data of a feature that is  compatible with the emission arising from a proto-pulsar wind nebula (PWN), the only hint for the existence of such elusive compact object is provided by the detection of hard emission in NuSTAR data up to $\sim$ 20 keV. We report on the simultaneous analysis of multi-epoch observations of \sna\ performed with \chandra, \xmm\ and \nustar. We also compare the observations with a state-of-the-art 3D magnetohydrodynamic (MHD) simulation of \sna. A heavily absorbed power-law, consistent with the emission from a PWN embedded in the heart of \sna, is needed to properly describe the high-energy part of the observed spectra. The spectral parameters of the best-fit power-law are in agreement with the previous estimate, and exclude diffusive shock acceleration as a possible mechanism responsible for the observed non-thermal emission. The information extracted from our analysis are used to infer the physical characteristics of the pulsar and the broad-band emission of its nebula, in agreement with the ALMA data. Analysis of the synthetic spectra also show that, in the near future, the main contribution to Fe K emission line will originate in the outermost shocked ejecta of \sna.

\end{abstract}

\section{Introduction} 
\label{sec:intro}

\sna, located at 51.4 kpc from Earth \citep{pan99}, in the Large Magellanic Cloud (LMC), was a core-collapse supernova (SN) discovered on February 23, 1987 \citep{wls87}. The dynamical evolution of the SN remnant (SNR) is strictly related to the highly inhomogenous circumstellar medium (CSM), made of a dense and clumpy ring-like structure within a diffuse HII region \citep{sck05}. The evolution of \sna\ has been extensively monitored in various wavelengths \citep{mcr93,mcf16}: in particular, the X-ray band is ideal to investigate the interaction of the shock front with the CSM (e.g. \citealt{bbm97,pzb06,hga06,mhs12,svc21,mhs21,rpz21})  and the emission of the expected central compact relic of the SN explosion.

Despite the excellent understanding granted with deep and continuous observations and the neutrinos detection from the SN (\citealt{bbb87}), which strongly indicate the formation of a neutron star (NS, \citealt{vis15}), a direct detection of the elusive compact object of \sna\ is still missing. The most likely explanation for this non-detection is ascribable to the absorption by the innermost ejecta, i.e. the dense and cold material ejected by the SN and surrounding the putative compact object \citep{fcr87}: because of the young age of \sna, these ejecta are still very dense and the reverse shock generated in the outer shell of the SNR has not heated them yet. Photo-electric absorption from this metal-rich material can hide the X-ray emission of a compact leftover \citep{omp15,alf18a,erl18,pbg20}. 

The X-ray emission from a young NS may include a significant non-thermal component: the synchrotron radiation arising from its magnetosphere or from the pulsar wind nebula (PWN) associated with the rotating NS . Recently, {\it ALMA} images showed a {\it blob} structure -- located at the position where the compact object is expected to be -- whose emission is compatible with the radio emission of a PWN \citep{cmg19}, though the same authors warned that the blob could be associated with other physical processes. \citet{gmo21} (hereafter G21) found indication for a PWN emitting in the hard X-ray band through the joint analysis of \chandra\ and \nustar\ data collected in 2012 and 2014, though they cannot completely exclude that the non-thermal emission might be due to diffusive shock acceleration (DSA) at the shock front of the SNR. On the contrary, \citet{alf21} (hereafter A21) favor a thermal origin, with kT $\sim 4$ keV, for this hard emission from the analysis of \xmm\ and \nustar\ data. 

In this paper, we re-analyzed data of \sna\ collected in 2012, 2014 and 2020 by \chandra, \xmm\ and \nustar\ with the aim to scrutinize the hypothesis of a PWN deeply embedded in \sna\ and to understand what is the origin of the different interpretation of the data discussed in G21 and A21. The analysis of the variability of the hard X-ray emission over 8 years (a significantly longer time lapse than 2 years, as in G21) can provide more conclusive indications on the nature of the source; for instance, it can help to discern between the DSA and PWN scenarios (see G21). Finally, in order to discriminate between the non-thermal and thermal origin of the emission above 10 keV as proposed by G21 and A21, respectively, we compared the observed spectra with those predicted by a state-of-the-art three-dimensional (3D) magneto-hydrodynamic (MHD) simulation of \sna\ (model B18.3) described by \citet{oon20} (hereafter Or20). This allowed us to provide a coeherent description of the multi-epoch emission from 0.5 to 20 keV based both on standard data analysis and comparison of the data with the MHD simulation. In a companion paper (Dohi et al, in preparation), we investigate the detectability of the thermal emission counterpart from the putative central compact object (CCO) in \sna.

The paper is organized as follows: in Sect.~\ref{sec:x-ray_data}, we present the data and describe their analysis either adopting a traditional approach or using the MHD model B18.3, and we discuss the possible properties of the putative PWN, using available data as constraints for the spectrum;
in Sect.~\ref{sect:disc}, we discuss the results and their implications; in Sect.~\ref{sec:sum}, we summarize the main findings and draw our conclusions;
in Appendix \ref{app:absorbed_spec} we investigate the detectability of the PWN in the soft X-ray band ($0.5-8$ keV).

\section{X-ray data analysis} \label{sec:x-ray_data}

We focused on the epochs when simultaneous (or almost simultaneous) \chandra, \xmm\ and \nustar\ observations of \sna\ were performed. In particular, we analyzed data collected with \chandra/ACIS-S, \xmm/pn, \xmm/RGS and \nustar/FPMA,B in 2012, 2014 and 2020. Details on the observations are reported in Table \ref{tab:obs}.

\begin{table*}[!ht]
  \centering
   \caption{Summary of the main characteristics of the analyzed observations.}
  \begin{tabular}{c|c|c|c|c}
   Telescope &OBS ID& PI& Date (yr/month/day) & Exposure time (ks) \\
   \hline\hline
   & 3830& Burrows& 2003/07/08 & 45\\
   & 13735& Burrows& 2012/03/28 &48\\
   & 14417& Burrows& 2012/04/01&27\\
   \chandra\ & 15809& Burrows & 2014/03/19 &70\\ 
   & 15810& Burrows& 2014/09/20 &48\\
   &  22425& Burrows & 2020/09/12 &  61\\
   & 24652 & Burrows &  2020/09/17&  29\\
   \hline
   & 40001014003& Harrison& 2012/09/08&136\\
   & 40001014004& Harrison& 2012/09/11&200\\
   & 40001014007& Harrison& 2012/10/21&200\\
   \nustar\ & 40001014018& Harrison& 2014/06/15& 200\\
   & 40001014020& Harrison& 2014/06/19&275\\
   & 40001014023& Harrison& 2014/08/01&428\\
   & 40501004002& Alp & 2020/05/13 & 183 \\
   & 40501004004& Alp & 2020/05/27 & 159 \\
   \hline
   &0690510101& Haberl & 2012/12/11 & 70/53$^a$ \\
   \xmm& 0743790101& Haberl & 2014/11/29 & 81/50$^a$ \\
   &0862920201& Haberl & 2020/11/24 & 81/50$^a$\\
  \end{tabular}
  
  $^a$ \xmm/pn Unfiltered/filtered exposure time
  \label{tab:obs}
\end{table*}

Spectral analysis has been performed with XSPEC (v12.11.1, \citealt{arn96}) in the $0.5-8$ keV, $0.3-10$ keV, $0.35-2.5$ keV and $3-20$ keV bands for the \chandra, \xmm/pn, \xmm/RGS and \nustar\ data, respectively. We excluded from the analysis \xmm/RGS bins with variance equal to 0 in order to avoid potential errors in the $\chi^2$ estimate.

We reprocessed \chandra, \xmm\ and \nustar\ data with the standard pipelines available within CIAO v4.12.2, SAS v17.00.00 and NuSTARDAS v2.1.1, respectively. We extracted \chandra\ and \nustar\ spectra following G21, and the \xmm\ spectra following \citet{svc21} (hereafter S21). All the spectra are rebinned optimally following the procedure described by \citet{kb16}, and the background spectrum for subtraction was extracted from a nearby region immediately outside of the source (see G21). We verified that our results are not affected by the choice of the background regions.

\subsection{Standard analysis of the spectra}
\label{sub:spectra_analysis}

We considered three clusters of observations, one for each epoch, namely 2012, 2014, and 2020.
We simultaneously analyzed \chandra, \xmm\ and \nustar\ spectra for each selected year by adopting a model composed by a foreground absorption component (\texttt{TBabs} model in XSPEC; \citealt{arn96}), and components of optically thin isothermal plasma in non-equilibrium of ionization (\texttt{vnei} model). We included a multiplicative factor $C_i$ which takes into account cross-calibration between different detectors (the index $i$ running through the various instruments). For the \xmm/RGS spectra, a \texttt{gsmooth} and a \texttt{vashift} components are also included, with values fixed to those of S21, to account for line broadening and line shift due to the systemic motion of \sna, respectively. The foreground column density N$_{\rm{H}}$ is fixed to $2.35 \times 10^{21} \, \mathrm{cm}^{-2}$ \citep{pzb06}. In the following, we refer to this model as \emph{2-kT model} or \emph{3-kT model}, depending on the number of \texttt{vnei} components included. We found differences from 1 of the cross-calibration factors $<5\%$ between \chandra\ and  \xmm\ spectra, and $<10\%$ between \nustar\ and \xmm\ (and between \nustar\ and \chandra), compatible with the characteristic values (\citealt{mhm15}). 

Temperature, emission measure, ionization parameter and abundances of O, Ne, Mg, Si, S and Fe were left free to vary in the fitting process and the other abundances were kept fixed to typical values for the LMC \citep{rd92}.

\begin{figure*}[!ht]
    \centering
    \begin{minipage}{0.48\textwidth}
    \includegraphics[angle=270,width=.95\textwidth]{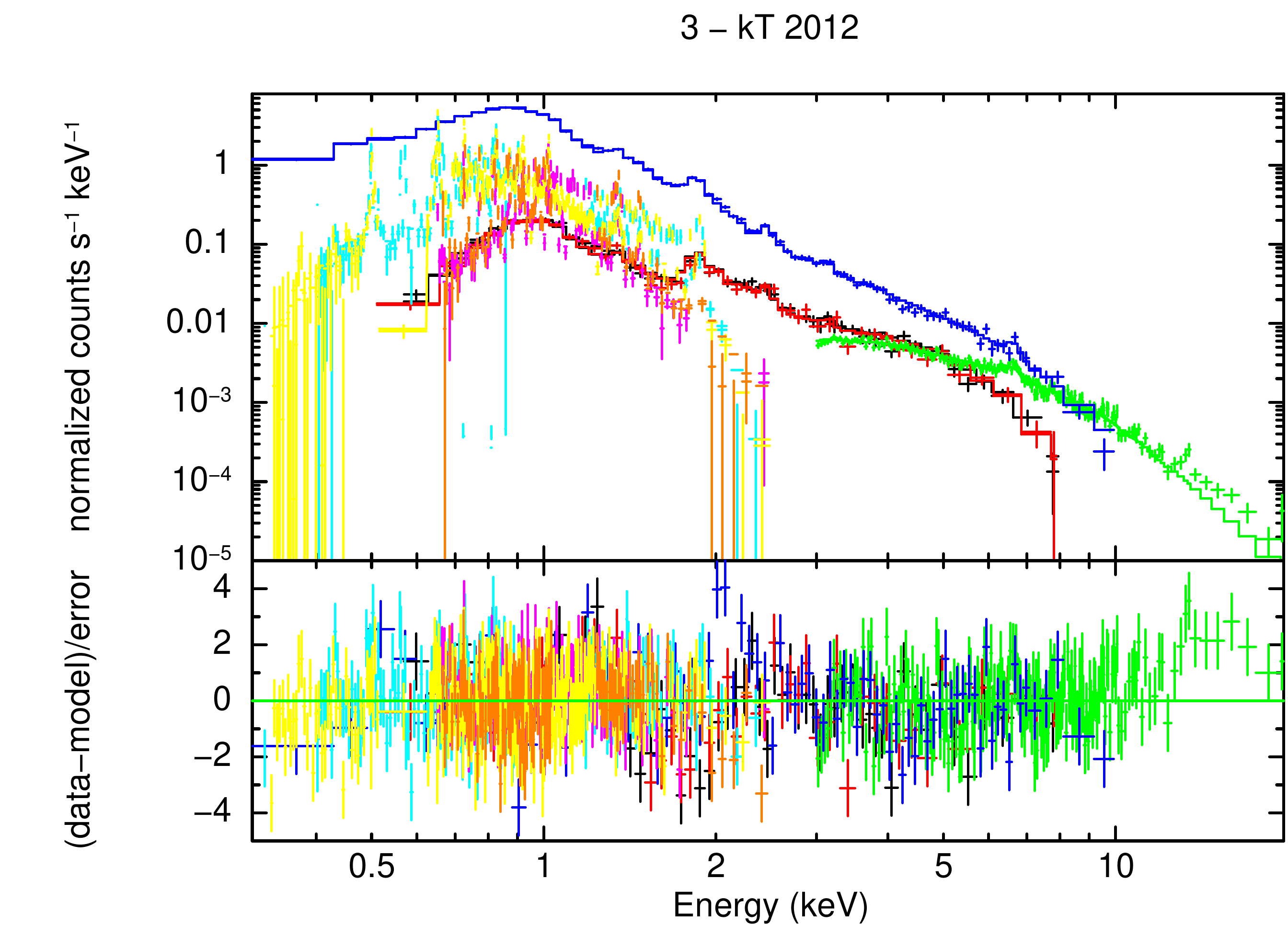}
    
    \end{minipage}
    \begin{minipage}{0.48\textwidth}
    \includegraphics[angle=270,width=.95\textwidth]{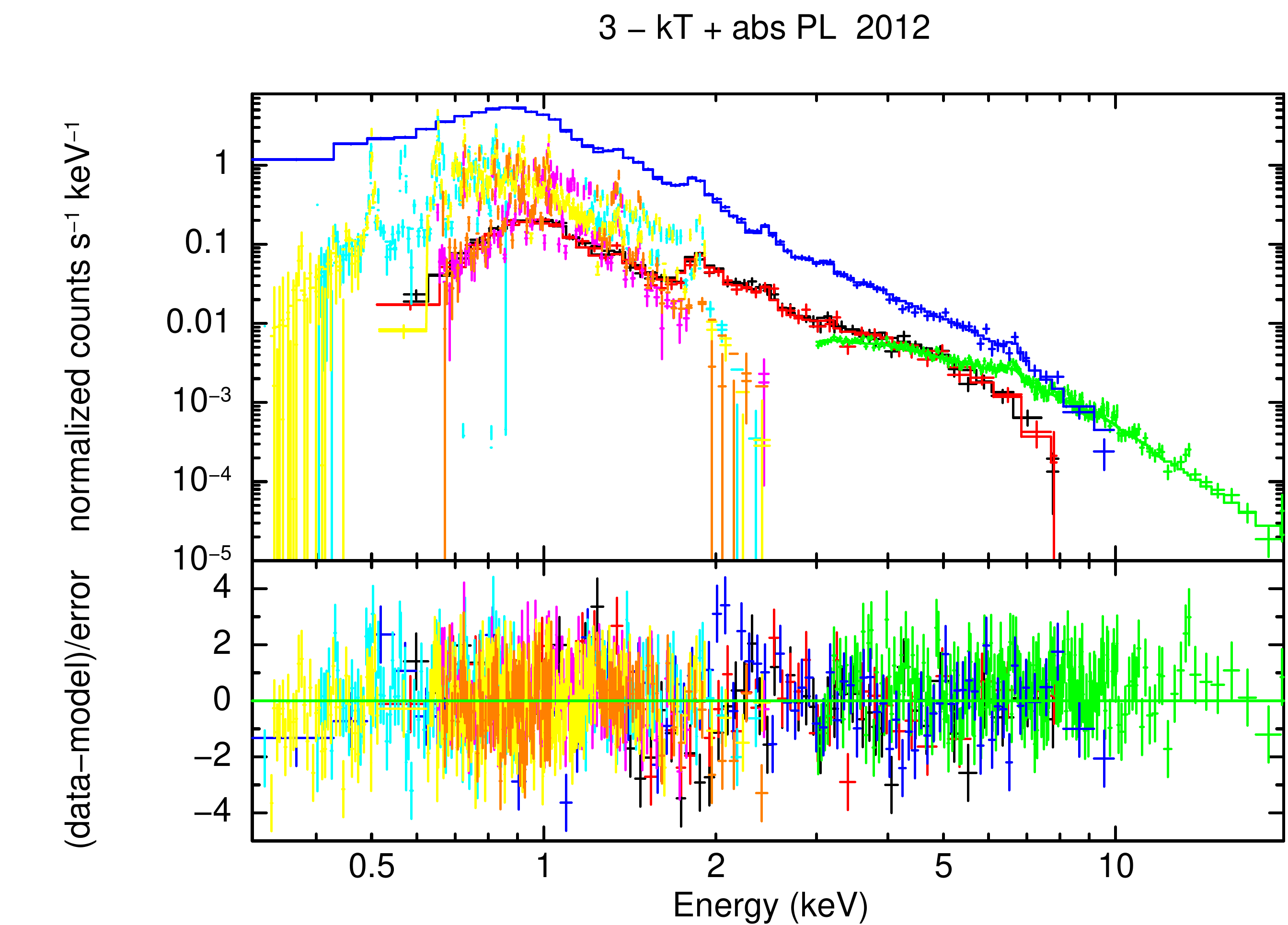}
    \end{minipage}
    \begin{minipage}{0.48\textwidth}
    \includegraphics[angle=270,width=.95\textwidth]{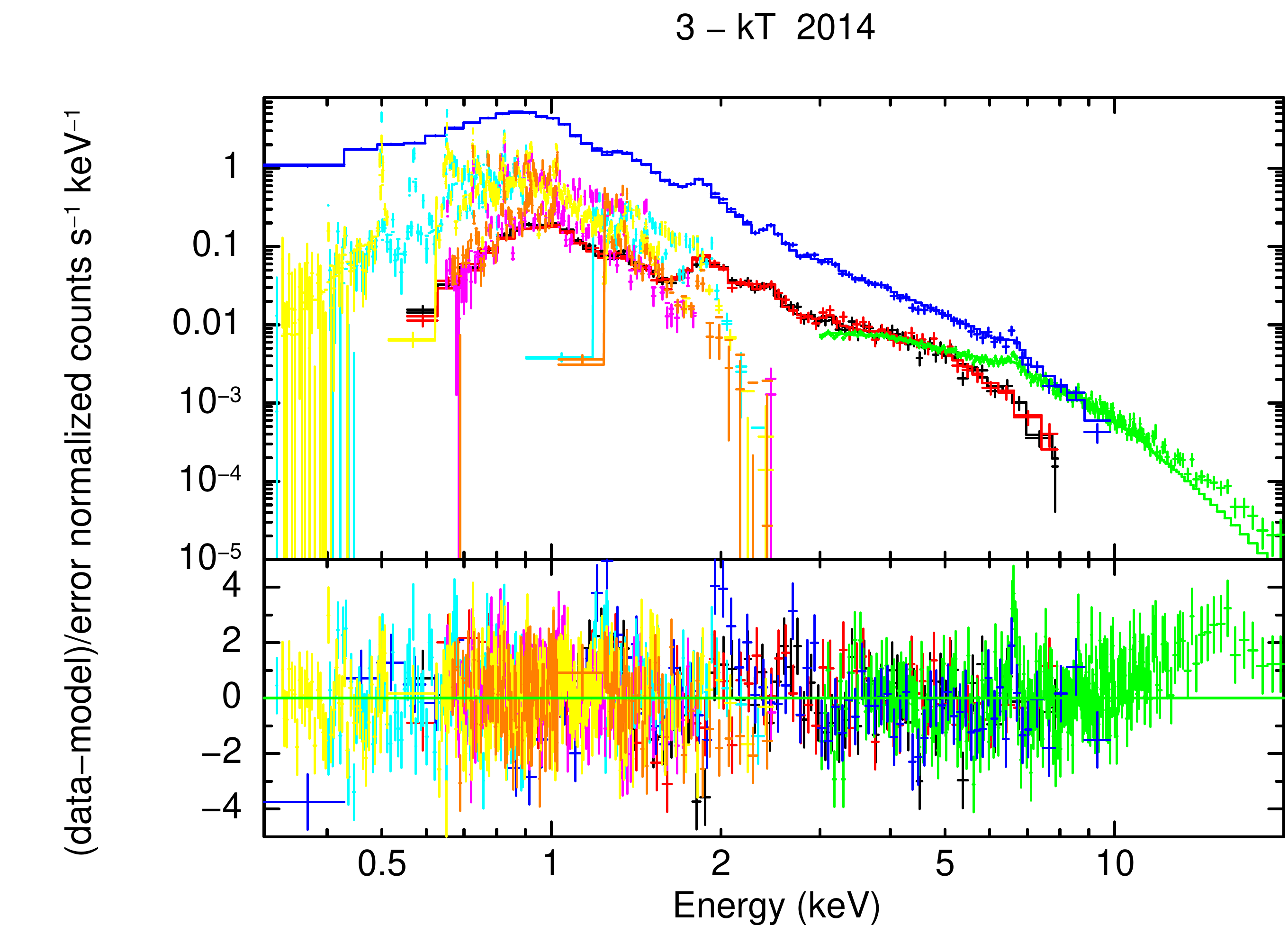}
    \end{minipage}
    \begin{minipage}{0.48\textwidth}
    \includegraphics[angle=270,width=.95\textwidth]{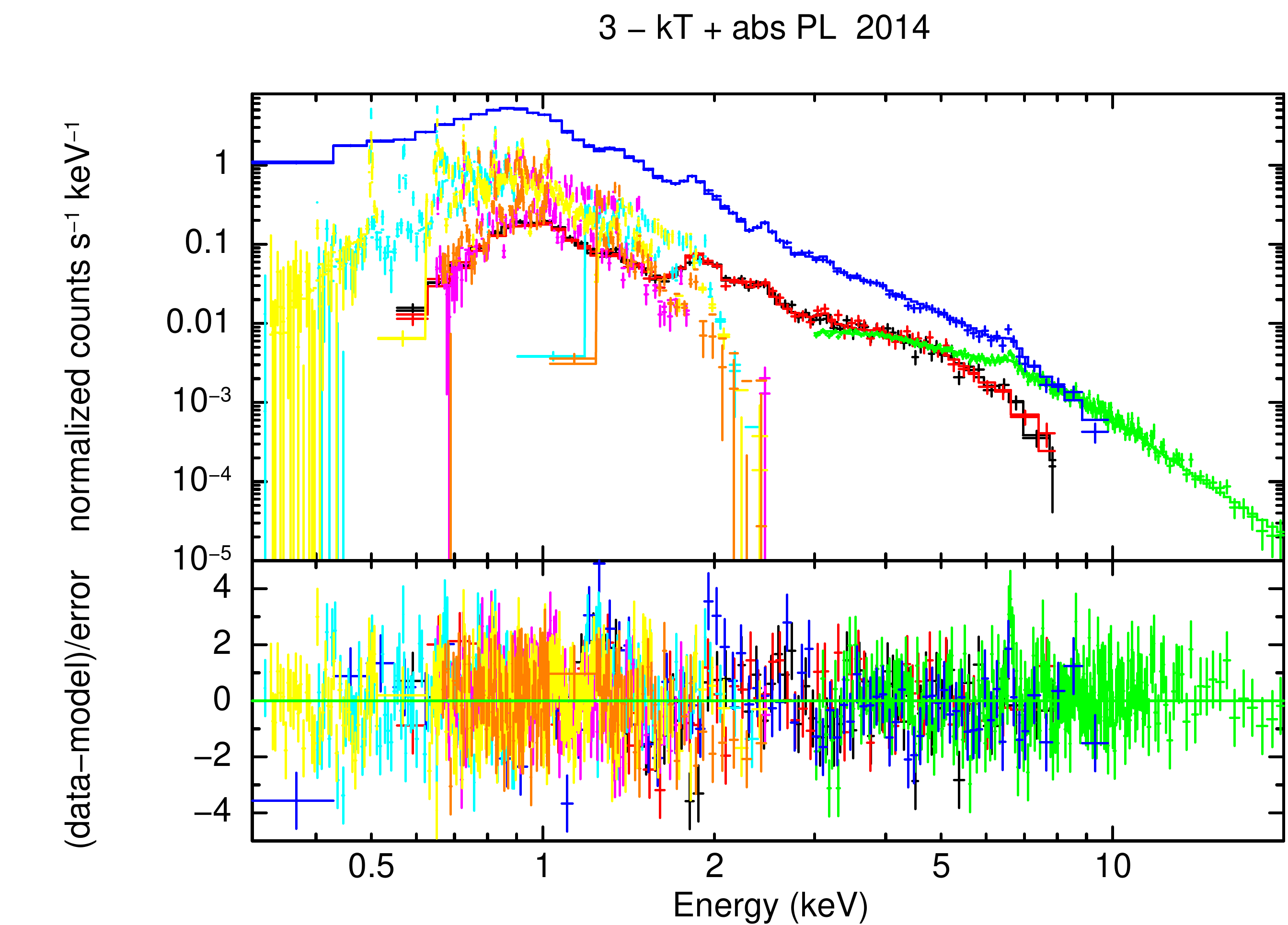}
    \end{minipage}
    \begin{minipage}{0.48\textwidth}
    \includegraphics[angle=270,width=.95\textwidth]{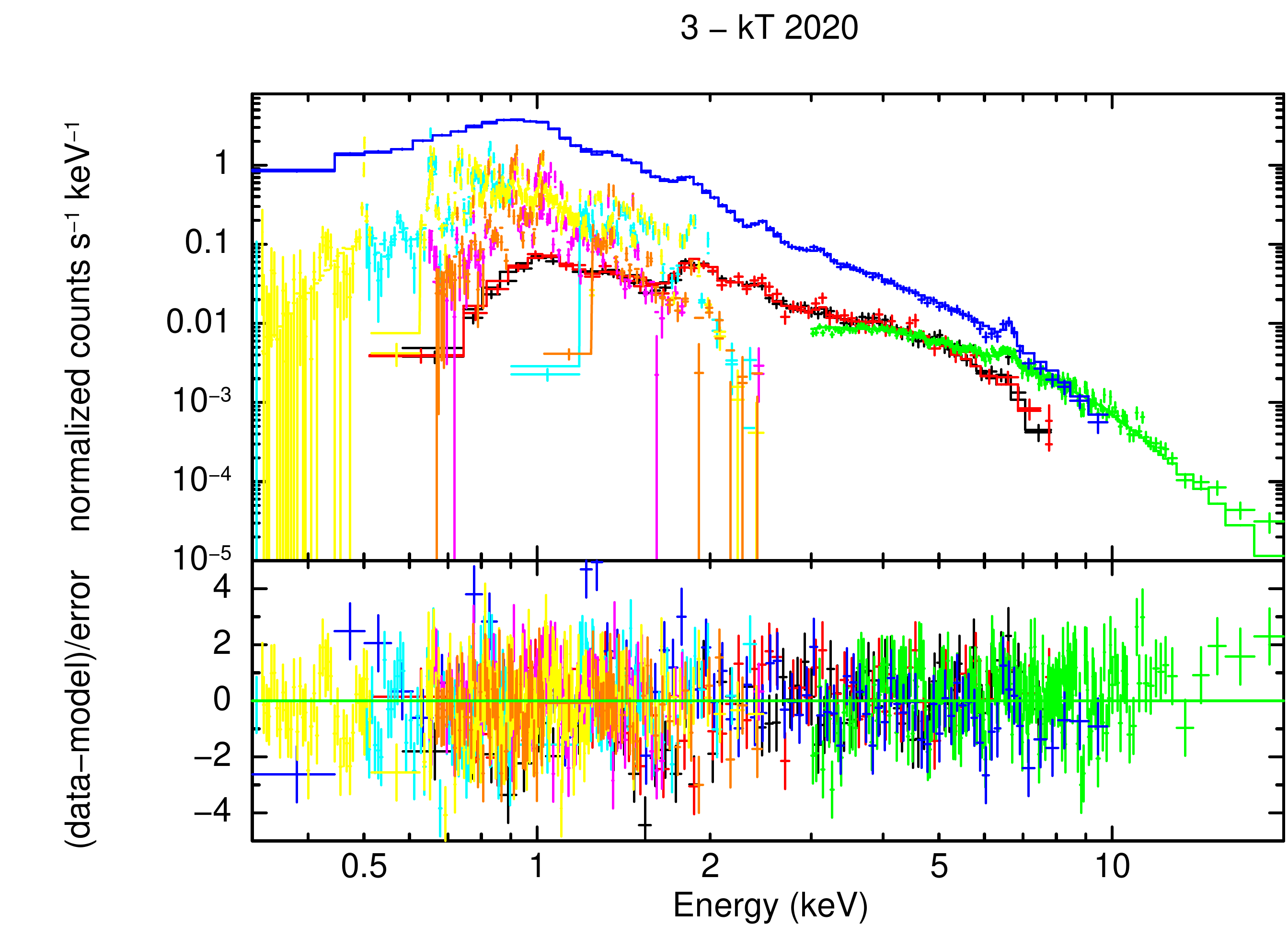}

    \end{minipage}
    \begin{minipage}{0.48\textwidth}
    \includegraphics[angle=270,width=.95\textwidth]{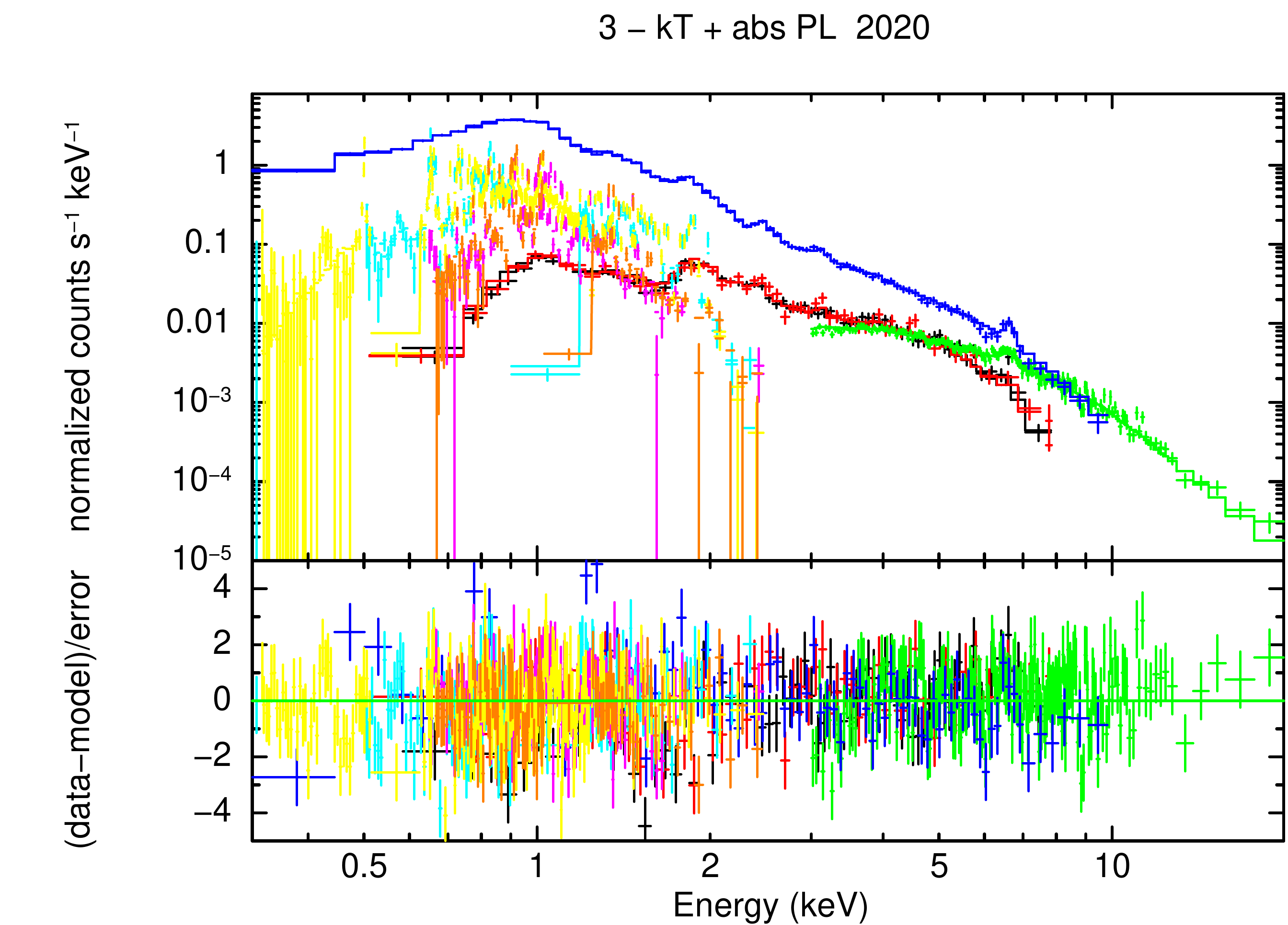}
    \end{minipage}
    \caption{\chandra/ACIS-S, \xmm/pn,RGS and \nustar/FPMA,B spectra of \sna\ in 2012 (first row), 2014 (second row) and 2020 (third row) with the corresponding residuals. \nustar\ spectra are summed for visualization purposes only. Different colors identify different spectra: \chandra\ in black and red; \nustar\ in green; \xmm/PN in blue; \xmm/RGS in yellow, cyan, orange and light green. The spectra are simultaneously fitted either with the \emph{3-kT model} (left panels) or with the \emph{3-kT model} plus an absorbed PL (right panels), representing the PWN scenario.}
    \label{fig:fenom}
\end{figure*}

Before fitting simultaneously the spectra collected by all telescopes, we analyzed the \chandra\ and \xmm\ spectra separately. We reproduced the same results as G21 for the 2012 and 2014 \chandra\ data and as S21 for the multi-epoch \xmm\ data. In particular, both \chandra\ and \xmm\ data can be well reproduced by considering the \emph{2-kT} model. Including a third thermal component does not improve the fit quality of the \chandra\ spectra, whilst it leads to a significantly better description of the \xmm\ data due to the higher statistics of the data and sensitivity of the \xmm\ detectors. Thus, we fitted the 0.5-20 keV spectra jointly collected by \chandra, \xmm, \nustar\ with the \emph{3-kT model}, since it provides the best description in this case. The spectra, with the corresponding best-fit \emph{3-kT model} for each epoch, are shown in the left panels of Fig. \ref{fig:fenom}. The best-fit parameters are shown in Table \ref{tab:fit_whole}.

\begin{table*}[!ht]
  \centering
  \caption{Best-fit parameters of the \emph{3-kT model} and 3-kT plus absorbed PL model}
 
  \begin{tabular}{c|c||c|c|c||c|c|c}
  \hline\hline
  \multicolumn{2}{c||}{ }&  \multicolumn{3}{c||}{3-kT model} & \multicolumn{3}{c}{3-kT + absorbed PL model}\\
  \hline
  Component & Parameter & 2012 & 2014 & 2020 & 2012 & 2014 & 2020\\
  \hline
  \texttt{TBabs} & N$_{\rm{H}}$ (10$^{22}$ cm$^{-2}$) & \multicolumn{6}{c}{0.235 (fixed)} \\
  \hline
  & kT$_1$ (keV) & 0.42$_{-0.03}^{+0.04}$& 0.41$_{-0.02}^{+0.03}$& 0.48$\pm 0.06$&0.40$\pm 0.03$ &0.40$\pm 0.03$& 0.48$_{-0.07}^{+0.05}$\\
  & O & 0.158$_{-0.008}^{+0.015}$& 0.19$_{-0.02}^{+0.01}$& 0.16$_{-0.01}^{+0.03}$&0.169$_{-0.013}^{+0.012}$ & 0.196$_{-0.016}^{+0.017}$&0.16$_{-0.02}^{+0.03}$\\
  & Ne& 0.42$_{-0.02}^{+0.04}$& 0.50$\pm 0.04$& 0.38$_{-0.04}^{+0.02}$&0.46$_{-0.02}^{+0.04}$ & 0.54$_{-0.03}^{+0.08}$&0.38$ \pm 0.5$\\
  & Mg& 0.56$_{-0.04}^{+0.06}$& 0.62$_{-0.06}^{+0.03}$& 0.49$_{-0.02}^{+0.04}$& 0.59$_{-0.04}^{+0.02}$& 0.64$\pm 0.4$&0.50$_{-0.04}^{+0.03}$\\
  \texttt{vnei$_1$} & Si & 0.82$_{-0.04}^{+0.05}$& 0.84$\pm 0.4$& 0.71$\pm 0.04$& 0.83$_{-0.05}^{+0.04}$&0.84$\pm 0.04$ &0.71$_{-0.04}^{+0.05}$\\
  & S & 0.67$_{-0.08}^{+0.07}$& 0.72$_{-0.07}^{+0.06}$& 0.57$_{-0.07}^{+0.06}$& 0.64$\pm 0.07$& 0.69$_{-0.06}^{+0.07}$&0.57$\pm 0.06$\\
  & Fe& 0.368$_{-0.016}^{+0.023}$& 0.396$_{-0.017}^{+0.018}$& 0.337$_{-0.011}^{+0.020}$& 0.398$_{-0.020}^{+0.016}$& 0.42$\pm 0.02$&0.342$_{-0.011}^{+0.022}$\\ 
  & $\tau_1$ (10$^{11}$ s/cm$^3$)& 1.02$_{-0.06}^{+0.17}$& 0.77$\pm 0.13$& 0.77$_{-0.05}^{+0.20}$& 1.04$_{-0.11}^{+0.15}$&0.78$_{-0.11}^{+0.20}$ &0.79$_{-0.12}^{+0.15}$\\
  & EM$_1$ ($10^{58}\, \mathrm{cm}^{-3}$) & 13.7$_{-1.1}^{+1.0}$& 8.3$\pm{1.0}$& 5.9$\pm 0.9$& 12.8$_{-1.1}^{+0.9}$ & 8.3$\pm 0.8$ & 5.9$_{-0.8}^{+0.8}$\\
  \hline
  & kT$_2$ (keV)& 0.87$_{-0.03}^{+0.05}$& 0.83$_{-0.02}^{+0.01}$& 0.96$_{-0.05}^{+0.06}$& 0.84$_{-0.03}^{+0.02}$& 0.827$_{-0.020}^{+0.012}$&0.96$_{-0.05}^{+0.03}$\\
  \texttt{vnei$_2$}& $\tau_2$ (10$^{11}$ s/cm$^3$)& 1.83$_{-0.17}^{+0.19}$& 2.11$_{-0.12}^{+0.24}$& 1.4$_{-0.2}^{+0.3}$& 1.7$_{-0.2}^{+0.4}$&1.9$_{-0.2}^{+0.3}$ &1.41$_{-0.10}^{+0.25}$\\
  & EM$_2$ ($10^{58}\, \mathrm{cm}^{-3}$) & 16.3$^{+0.9}_{-0.6}$& 17.3$_{-1.1}^{+0.6}$& 11.8$_{-1.0}^{+0.9}$ &  15.1$_{-1.0}^{+0.9}$ & 15.8$_{-0.4}^{+0.9}$ & 11.6$_{-0.6}^{+0.5}$\\
  \hline
  & kT$_3$ (keV) & 3.54$\pm 0.07$& 3.36$_{-0.05}^{+0.12}$& 3.34$_{-0.07}^{+0.08}$&2.91$_{-0.10}^{+0.07}$ & 2.87$_{-0.14}^{+0.04}$&3.18$_{-0.20}^{+0.012}$\\
  \texttt{vnei$_3$} & $\tau_3$ (10$^{11}$ s/cm$^3$) & 0.89$_{-0.05}^{+0.08}$& 1.01$\pm 0.5$ & 1.13$_{-0.09}^{+0.10}$&1.09$_{-0.1}^{+0.14}$ &1.28$_{-0.11}^{+0.12}$ & 1.21$_{-0.13}^{+0.15}$\\
  & EM$_3$ ($10^{58}\, \mathrm{cm}^{-3}$) & 5.51$_{-0.08}^{+0.11}$&6.86$_{-0.18}^{+0.09}$& 9.2$_{-0.3}^{+0.4}$&  6.8$_{-0.2}^{+0.5}$& 8.0 $\pm 0.3$ & 9.5$_{-0.4}^{+0.5}$\\
  \hline
  & $\Gamma$ & /& /& /& 3.0$_{-0.2}^{+0.4}$& 2.9$_{-0.4}^{+0.3}$ & 2.6 $_{-1.8}^{+0.7}$ $^b$\\
 \texttt{power-law}$^{a}$ &norm ($10^{-4}$ ph/s/keV/cm$^2$) & /& /& /&15$_{-5}^{+25}$ & 10$^{+14}_{-9}$& 2$_{-1}^{+9}$ $^b$ \\
 \hline
 \multicolumn{2}{c||}{Flux$_{0.5-8}$ (10$^{-13}$ erg/s/cm$^2$)} &96.1$\pm 0.05 $ &98.3$_{-0.4}^{+0.5}$ &91.3$_{-0.6}^{+0.5}$& '' & '' & ''\\
 \multicolumn{2}{c||}{Flux$_{10-20}$ (10$^{-13}$ erg/s/cm$^2$)} &1.19$\pm 0.09 $ &1.28$\pm 0.06$ &1.52$_{-0.14}^{+0.13}$ & 1.51$_{-0.11}^{+0.13}$&1.59$_{-0.08}^{+0.09}$ & 1.76$_{-0.17}^{+0.15}$\\
 \hline
 \multicolumn{2}{c||}{$\chi^2$/d.o.f.} & 2754/2139& 2912/2199& 2382/1897& 2680/2137& 2814/2197& 2377/1895\\
 \multicolumn{2}{c||}{$\chi^2_{red}$}& 1.29& 1.32& 1.26& 1.25& 1.28& 1.25\\
 \hline\hline
 \multicolumn{2}{c||}{ } & \multicolumn{3}{c||}{MHD model} & \multicolumn{3}{c}{MHD + absorbed PL model} \\
 \hline
 \multicolumn{2}{c||}{$\chi^2$/d.o.f. (RGS1 1st order)} & 893/457 &959/456 & 1677/434 &  / & / & /\\ 
 \multicolumn{2}{c||}{$\chi^2$/d.o.f. (pn, ACIS-S, FPMA,B)} & 1375/553& 1547/588& 1933/459& 864/551 & 694/586 & 1643/459\\ 
 \hline
 \end{tabular}
 
  \textbf{Notes.} Abundance values expressed with respect to \citet{wam00}. Abundance of other elements are kept fixed to those found by \citet{rd92}. Uncertainties are estimated at 90\% confidence level, unless otherwise stated. $^{a}$ The power-law  is coupled to the \texttt{vphabs} component built from the MHD model. $^{b}$ Uncertainties estimated at 68\% confidence level.
  \label{tab:fit_whole}
\end{table*}

 The resulting 3-kT best-fit model provides an acceptable description of the data in the 0.3-10 keV band in all epochs, even though clear residuals systematically show up at energies above 10 keV. Our best-fit values of temperatures are consistent at a 90\% confidence level with the results by A21. 

We also reproduced the results by A21, based on a  3-\texttt{vpshock} components model, and we noticed that the observed excess in the 10-20 keV band is not significant adopting this model. However, the measured values of the ionization ages are much higher than expected for \sna, as also noted by A21 (see also S21 and \citealt{rpz21}). In light of this, we discuss the results based on the \texttt{vnei} model, which indicates that all the plasma components are under-ionized, as also reported by, e.g., \citet{zmd09, pzb06, mob19, rpz21}; S21. 

As a next step, following G21, we investigated whether a power-law (PL) component, heavily absorbed by the inner cold ejecta (T$_{\rm{ej}} \sim 10$ K), added to the \emph{3-kT model} significantly improves the fit. The PL component is associated with the non-thermal emission arising from a PWN embedded in the center of \sna.

The absorption of cold ejecta was calculated from model B18.3 (Or20), by adopting the same approach as in G21: we used the XSPEC photoelectric absorption model \texttt{vphabs} and the distributions of temperature, density and abundance derived in the whole spatial domain in model B18.3 to properly calculate, for each epoch, the absorption by the ejecta for a source located where model B18.3 predicts the NS is, considering a kick velocity of 500 km/s\footnote{The absorption models for all the epochs have been made available in a public repository, as well the spectra and the best-fit models discussed in the whole paper, at the link 10.5281/zenodo.6006289}. We verified that our results do not change significantly if we consider kick velocities between 300 and 700 km/s. We estimated that the corresponding column density $\rm{N_H}$ is $\gtrsim 10^{23} \, \rm{cm}^{-2}$, two orders of magnitude higher than the foreground absorption.

Including the additional absorption component by the innermost ejecta is important to properly understand the physical origin of a putative non-thermal component: the emission stemming from the PWN is absorbed by this cloud of cold ejecta, leading to a negligible emission contribution at energies $\lesssim 5$ keV (see G21 and also Appendix \ref{app:absorbed_spec}). Indeed, we found that \emph{the additional absorbed PL component significantly improves the fit of the spectra at all epochs}. We verified that this additional component is statistically needed: its normalization is higher than 0 at the 5$\sigma$ confidence level in 2012 and 2014, and at the 2$\sigma$ confidence level in 2020. The improved $\chi^2$ values obtained when including the absorbed PL are $\Delta\chi^2_{2012} = -74; \Delta\chi^2_{2014} = -98; \Delta\chi^2_{2020} = -5$  with respect to the 3-kT scenario alone. 
The spectra fitted with this 3-kT plus absorbed PL model and the corresponding residuals are shown in the right panels of Fig. \ref{fig:fenom}. The best-fit parameters are reported in Table \ref{tab:fit_whole}.

 By looking at the residuals in Fig. \ref{fig:fenom}, we see that the 3-kT and 3-kT plus absorbed PL models provide a very similar description of the spectra in the 0.3-10 keV energy band, but residuals between 10 keV and 20 keV show up when the absorbed PL component is not included, indicating that the pure-thermal model is not the best suited for these data points. By adding the absorbed PL, these points are satisfyingly reproduced.

The PL parameters are better constrained by 2014 data, thanks to the higher statistics available and the smaller error bars, with $\Gamma$ spanning in the range [2.5-3.2]. For each year considered, the photon index $\Gamma$ and the normalization of the PL are consistent with being constant, and so it is also the flux in the 10-20 keV band. 
We checked that no significant changes in the description of the spectra are observed by fixing the photon index  to 2.8 (the average of the best-fit value derived through the epochs), with a variation $\Delta\chi^2\lesssim 2$ points.

We also noticed a decrease in the best-fit flux of the power-law for the 2020 data. 
However, we found no significant worsening of the fit quality when imposing the 2014 PL component to fit the 2020 dataset: we argue that the observed decrease in the flux derived for the 2020 data is not due to an intrinsic decrease of the non-thermal emission but it reflects the increase of the contribution from thermal emission in this band. The less relevant contribution of the PL component to the total X-ray spectrum in 2020 (whilst this relative contribution is significant in 2012 and 2014) is consistent with the scenario of non-thermal emission originating from a PWN: since we expect no significant variations of the PWN luminosity over a time-lapse of a decade \citep{Torres:2014}, the constant contribution of the PL component to the spectrum is gradually overwhelmed by the increasing contribution of the thermal emission. We will further discuss this aspect in Sect. \ref{sec:mhdmodel} and Sect. \ref{sect:disc}. 

To further investigate this aspect, we performed a multi-epoch simultaneous fit of all the data considered in this work, analogously to what done by G21. We simultaneously fit the 2012, 2014 and 2020 data  with a 3-\texttt{vnei} plus absorbed PL model with the normalization, electron temperature and ionization ages of the \texttt{vnei} components free to vary, while the abundance values are free to vary, but forced to be constant in time.

Coherently with what discussed above, we also fixed the photon index $\Gamma=2.8$, leaving only the PL normalization free to vary. We found a very significant improvement of the fit when including the absorbed PL in the model: from $\chi^2_{\rm{thermal}} = 7997$ (6163 dof) to $\chi^2_{\rm{PWN}} = 7831$ (6162 dof). Best-fit parameters of the simultaneous multi-epoch analysis are shown in Table \ref{tab:fit_multiepoch}\footnote{The best-fit parameters of the \emph{3-kT model} are not shown since these are just the same as those found with the single year analysis}. 
\begin{table*}[!ht]
  \centering
  \caption{Best-fit parameters of the multi-epoch 3-kT plus absorbed PL model }
 
  \begin{tabular}{c|c||c|c|c}
  \hline\hline

  Component & Parameter &  2012 & 2014 & 2020\\
  \hline
  \texttt{TBabs} & N$_{\rm{H}}$ (10$^{22}$ cm$^{-2}$) & \multicolumn{3}{c}{0.235 (fixed)} \\
  \hline
  & kT$_1$ (keV) & 0.40$\pm 0.02$ & 0.42$_{-0.04}^{+0.03}$ & 0.47$_{-0.06}^{+0.04}$ \\
  & O &  \multicolumn{3}{c}{0.176$_{-0.005}^{+0.006}$}\\
  & Ne&  \multicolumn{3}{c}{0.47$_{-0.02}^{+0.03}$}\\
  & Mg&  \multicolumn{3}{c}{0.59$\pm 0.02$}\\
  \texttt{vnei$_1$} & Si& \multicolumn{3}{c}{0.80$_{-0.02}^{+0.03}$}\\
  & S &  \multicolumn{3}{c}{0.64$_{-0.04}^{+0.04}$}\\
  & Fe&  \multicolumn{3}{c}{0.391$_{-0.008}^{+0.007}$}\\ 
  & $\tau_1$ (10$^{11}$ s/cm$^3$)& 1.03$_{-0.10}^{+0.19}$ & 0.80$\pm 0.11$ & 0.766$_{-0.010}^{+0.017}$\\
  & EM$_1$ ($10^{58}\, \mathrm{cm}^{-3}$) & $12.5_{-0.5}^{+0.7}$ & 8.8$_{-0.9}^{+0.4}$ & 5.7$_{-0.5}^{+0.4}$\\
  \hline
  & kT$_2$ (keV)& 0.827$_{-0.019}^{+0.010}$ & 0.828$_{-0.007}^{+0.008}$&0.98$_{-0.02}^{+0.03}$\\
  \texttt{vnei$_2$}& $\tau_2$ (10$^{11}$ s/cm$^3$)& 1.89$\pm 0.17$ &1.83$_{-0.13}^{+0.15}$ & 1.47$_{-0.17}^{+0.10}$\\
  & EM$_2$ ($10^{58}\, \mathrm{cm}^{-3}$) &15.8$_{-0.7}^{+0.2}$  &16.5$_{-0.4}^{+0.5}$ & 10.4$\pm 0.3$\\
  \hline
  & kT$_3$ (keV) & 2.96$\pm 0.06$ & 2.92$_{-0.05}^{+0.07}$&2.90$_{-0.03}^{+0.07}$\\
  \texttt{vnei$_3$} & $\tau_1$ (10$^{11}$ s/cm$^3$)& 1.05$_{-0.05}^{+0.08}$ &1.17$_{-0.08}^{+0.06}$ & 1.60$_{-0.11}^{+0.08}$\\
  & EM$_3$ ($10^{58}\, \mathrm{cm}^{-3}$) & 6.7$\pm 0.2$ & 7.8$\pm 0.2$ &  10.1$^{+0.3}_{-0.2}$\\
  \hline
  & $\Gamma$ & \multicolumn{3}{c}{2.8 (fixed)}\\
 \texttt{power-law}$^{a}$ &norm ($10^{-4}$ ph/s/keV/cm$^2$) & \multicolumn{3}{c}{7.1$_{-0.9}^{+0.6}$}\\
 & L$^{\rm{PWN}}_{10-20}$ (10$^{34}$ erg/s)& \multicolumn{3}{c}{3.0$\pm 0.3$} \\
 \hline
\multicolumn{2}{c}{ $\chi^2$ (d.o.f.) }& \multicolumn{3}{c}{7831 (6162)} \\
 \end{tabular}
 
  \textbf{Notes.} Abundance values expressed with respect to \citet{wam00} and assumed constant along the three epochs. Abundance of other elements are kept fixed to those found by \citet{rd92}. Uncertainties are estimated at 90\% confidence level. $^{a}$ The power-law  is coupled to the \texttt{vphabs} component built from the MHD model.
  \label{tab:fit_multiepoch}
\end{table*}
These confirm that the spectral characteristics of the absorbed PL found in this work are consistent with the findings by G21 and that the intrinsic PWN luminosity in the 10-20 keV band is constant over the three epochs at a value of $\sim 3 \times 10^{34}$ erg/s. This is also in agreement with the almost constant X-ray flux in the [$10-24$]~keV band reported by A21. 

G21 noted that the non-thermal emission revealed in the hard X-ray spectra of \sna\ could also be due to DSA at the SNR shock front, and that it is not possible to discern between DSA and PWN scenarios with 2012 and 2014 data alone.
Thanks to the data collected in 2020, we can now verify which of the two scenarios best describes the multi-epoch spectra. 
To this end, we considered the case in which the non-thermal emission originates from DSA occurring at the outer shells of \sna. 
Thus, we fitted the 2012, 2014 and 2020 data with a 3-kT plus PL model, namely without considering absorption from the ejecta, since now the source is not embedded in the cold and dense innermost ejecta but close to the remnant border.
The best-fit value of $\Gamma$ is $\sim 2$, compatible with the canonical photon index for synchrotron radiation. However, we found $\chi^2$ values worse than those obtained in the 3-kT plus absorbed PL scenario, with $\Delta\chi^2_{2012} =15; \Delta\chi^2_{2014} = 15; \Delta\chi^2_{2020} = 2$, with the same degrees of freedom. We further discuss this scenario in Sect. \ref{sect:disc}.

\subsection{Interpretation of the data with the MHD model} 
\label{sec:mhdmodel}

The analysis of \chandra, \xmm\ and \nustar\ data of \sna\ discussed in the previous section points towards a detectable non-thermal component, strongly absorbed and distinctly contributing to the hard X-rays. 
On the other hand, results by A21 based on 3-\texttt{vpshock} model points towards a thermal origin of this emission, though the ionization ages of the three components are not compatible with those expected for \sna\ (e.g., \citealt{zmd09, pzb06, mob19, rpz21}; S21). To further investigate this issue, we made a step beyond the standard fits of X-ray spectra with a few phenomenological spectral components, and used the 3D MHD model B18.3 by Or20 to investigate the thermal contribution of the shock-heated plasma to the X-ray spectra of \sna.

The advantage of this approach is to describe the multi-epoch thermal X-ray emission of \sna\ with a single \emph{physical} model which follows self-consistently the SNR evolution since the SN event, providing an almost continuous distribution of plasma temperatures, ionization parameters and emission measures at all epochs. Model B18.3 has been well constrained by multi-wavelength observations of the progenitor star, the parent SN and the SNR. It considers a stellar model that describes the main characteristics of the progenitor star of \sna\ \citep{2018MNRAS.473L.101U} and reproduces the main properties of the anisotropic SN explosion (\citealt{onf20}) and the thermal X-ray emission of the SNR in the 0.5-10 keV band (Or20).

Model B18.3 does not include a description of a hypothetical PWN. Thus, from the model, we only synthesized the thermal X-ray emission of the shock-heated plasma, assuming the abundances found by S21 and considering the effects of thermal and Doppler broadening of emission lines (derived from the distributions of ion temperatures and plasma velocities along the line of sight from model B18.3) in the synthesis of the spectra (see \citealt{mob19}). The plasma temperature, the ionization parameter, and the emission measure are provided by model B18.3 at all epochs in a self-consistent fashion.

A few ($< 10$) numerical cells of the model, comprising cold ($T < 10^6$~K) and dense ($n > 7.5 \times 10^4$~cm$^{-3}$) plasma (which contributes mainly to the very soft X-ray emission below 1 keV, which is not the focus of this work), were rejected from the analysis. In fact these cells correspond to dense clumps of plasma subject to a strong radiative cooling with a size comparable to (or smaller than) that of the cells. As a result, these clumps are not well described by the adopted spatial resolution.

\begin{figure*}[!ht]
     \begin{minipage}{0.32\textwidth}
    \includegraphics[width=\textwidth]{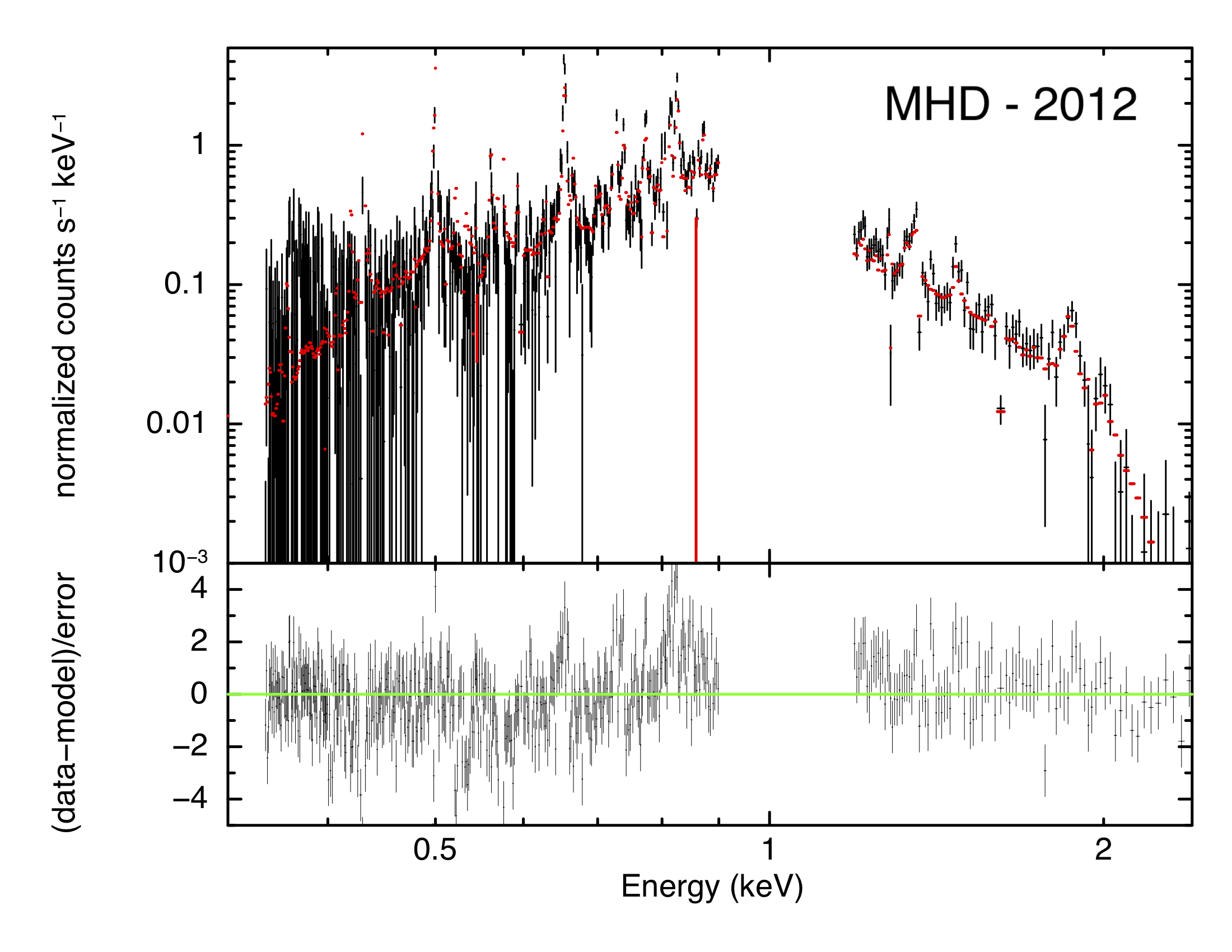}
    \end{minipage}
    \hfill
    \begin{minipage}{0.32\textwidth}
    \includegraphics[width=\textwidth]{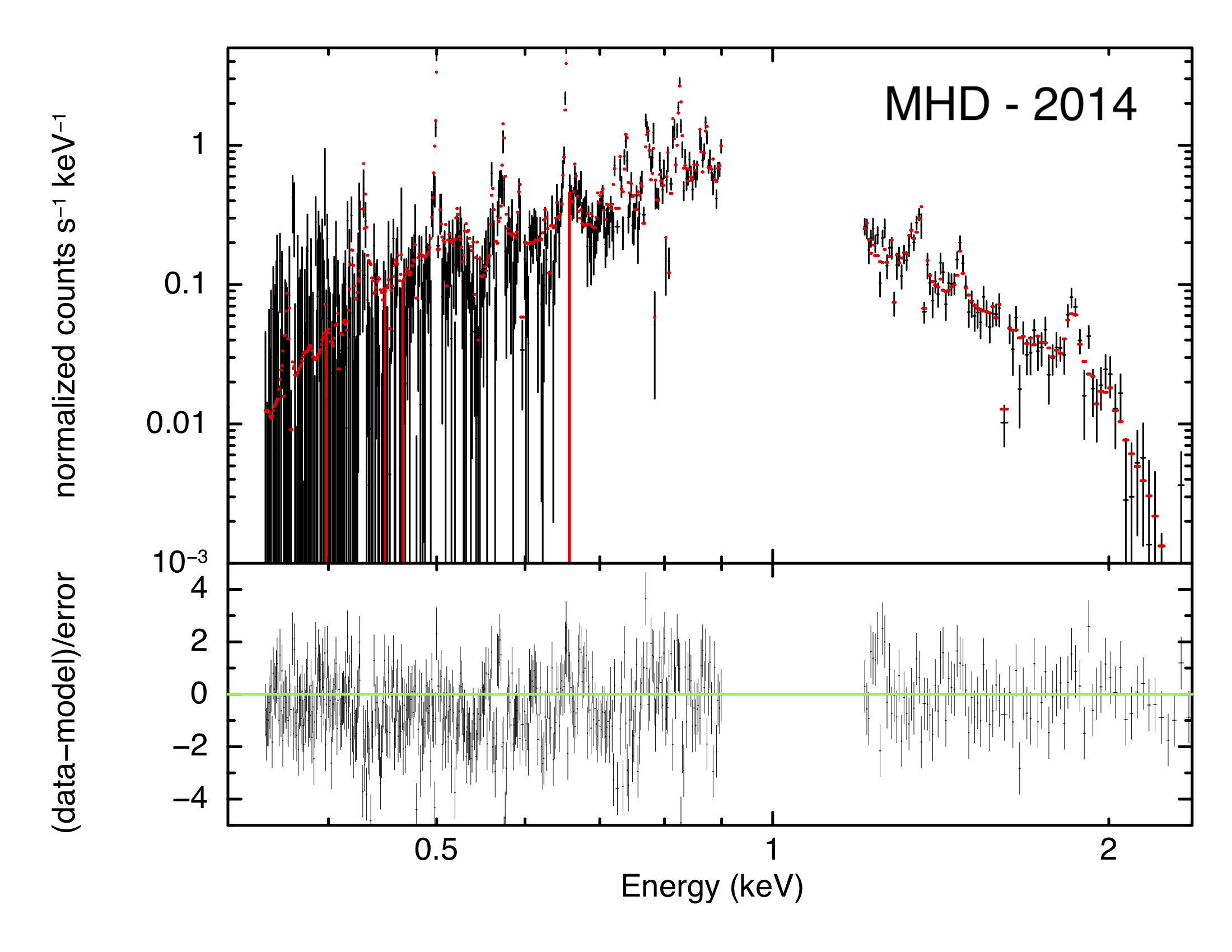}
    \end{minipage}
    \hfill
    \begin{minipage}{0.32\textwidth}
    \includegraphics[width=\textwidth]{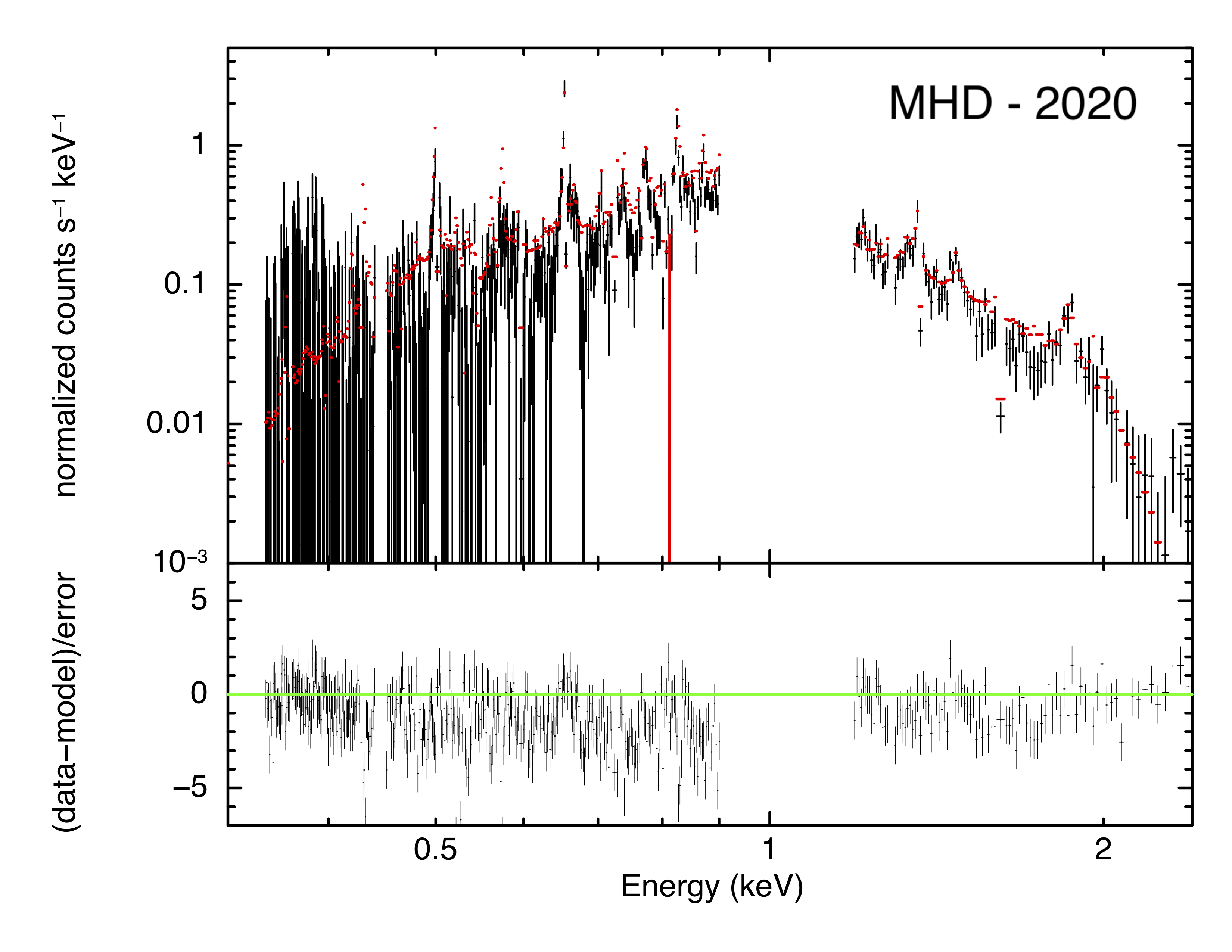}
    \end{minipage}
    \centering
    \begin{minipage}{0.32\textwidth}
    \includegraphics[angle=270,width=\textwidth]{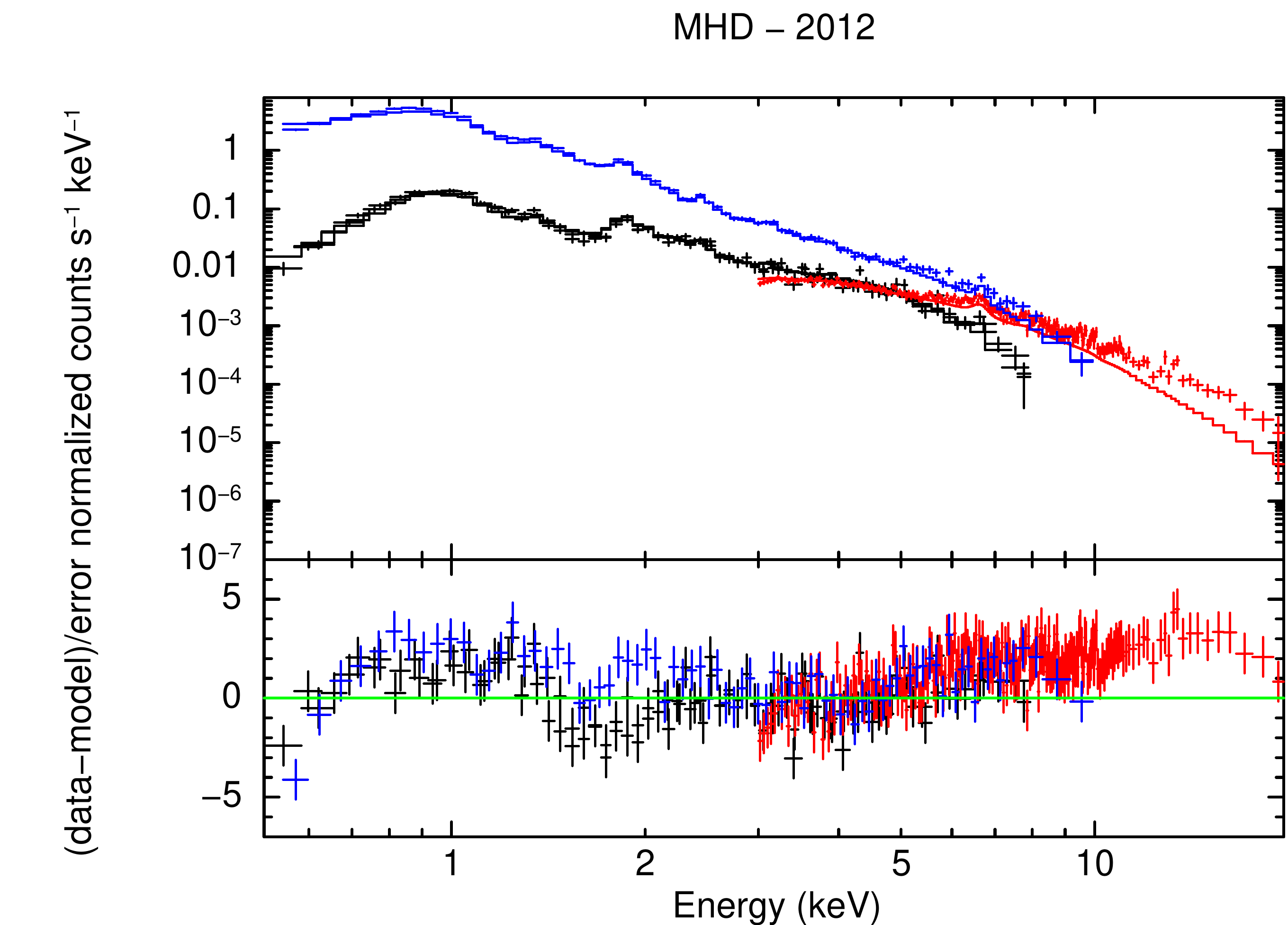}
    \end{minipage}
    \hfill
    \begin{minipage}{0.32\textwidth}
    \includegraphics[angle=270,width=\textwidth]{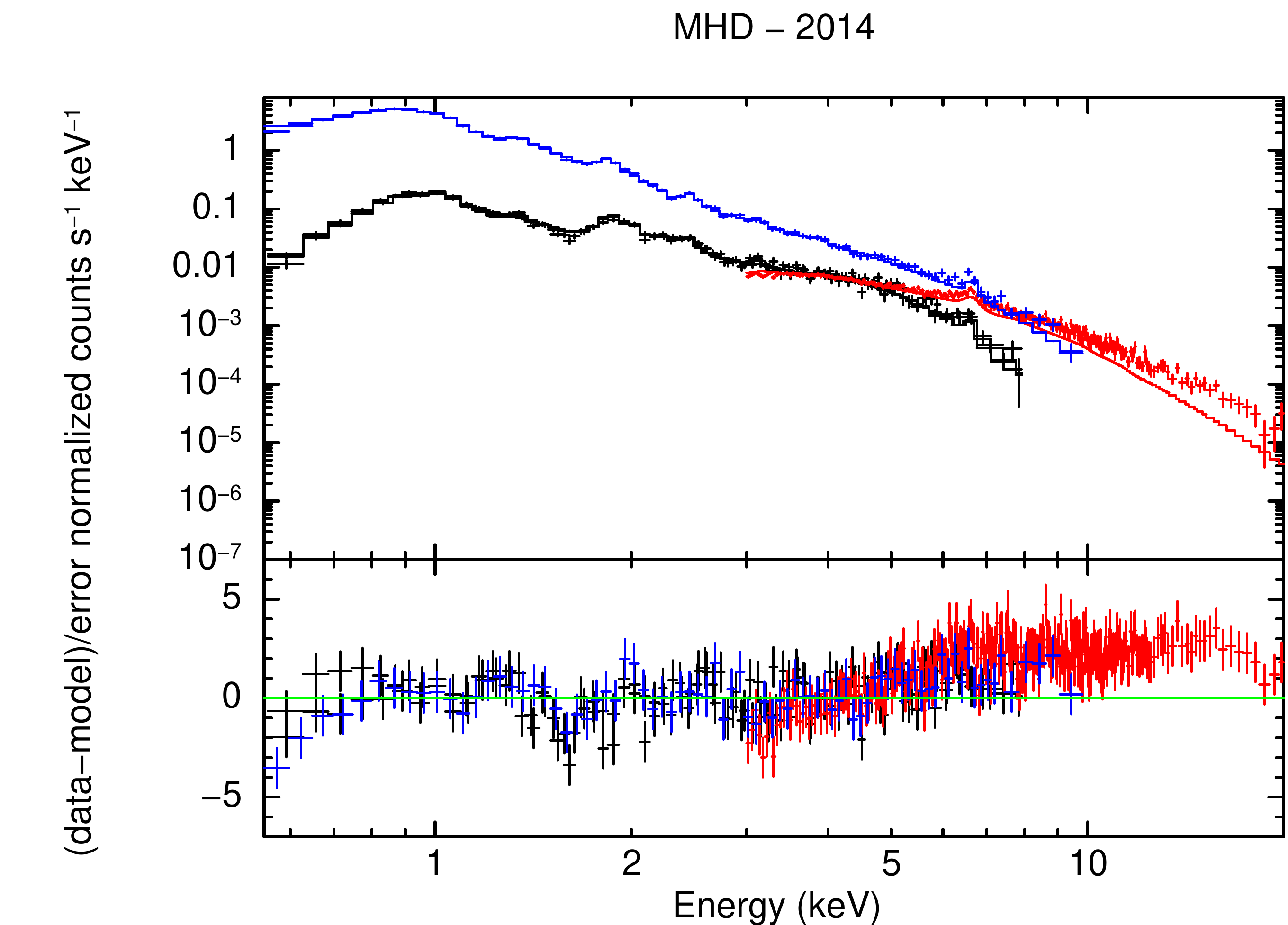}
    \end{minipage}
    \hfill
    \begin{minipage}{0.32\textwidth}
    \includegraphics[angle=270,width=\textwidth]{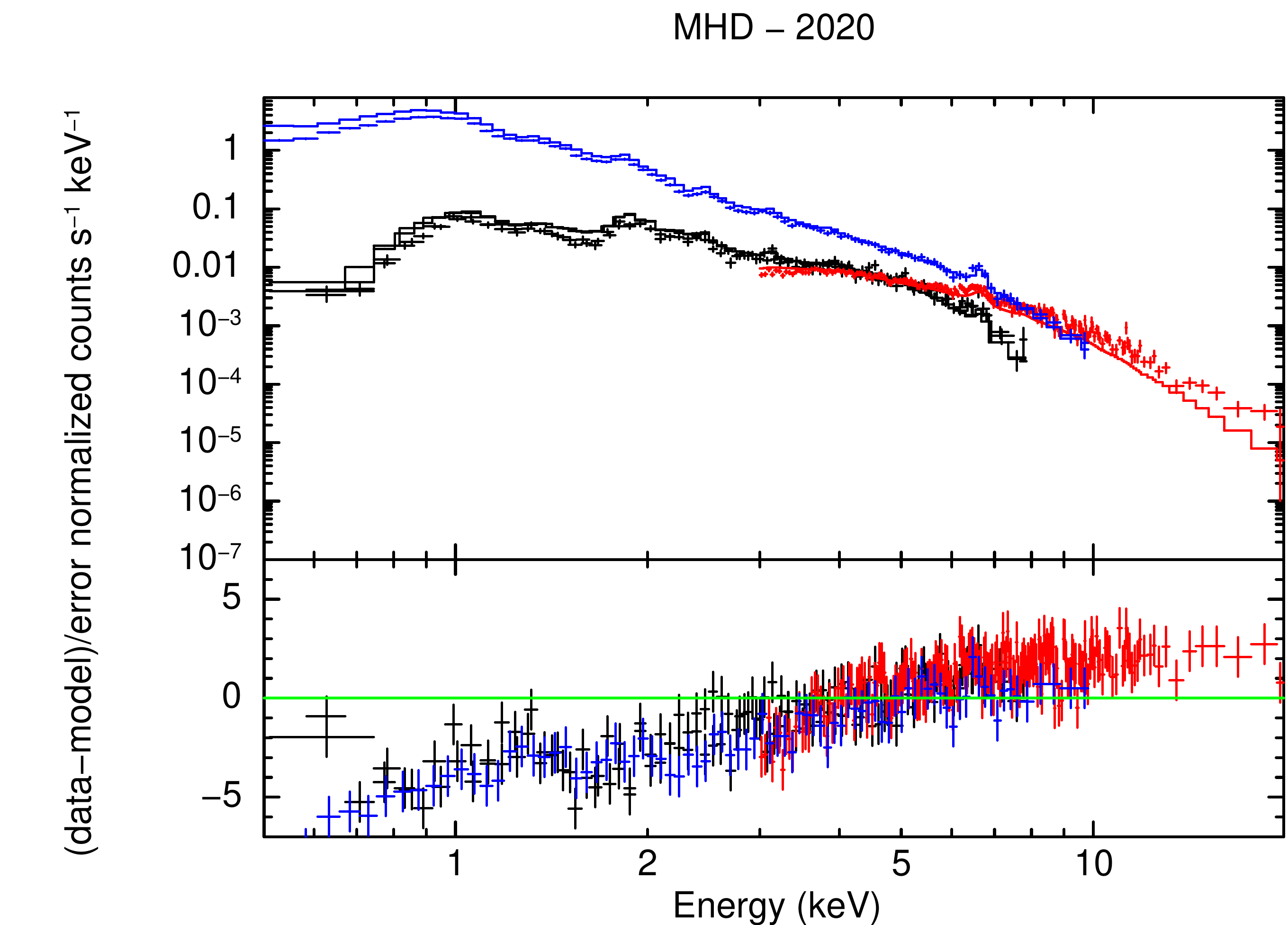}
    \end{minipage}
    \begin{minipage}{0.32\textwidth}
    \includegraphics[angle=270,width=\textwidth]{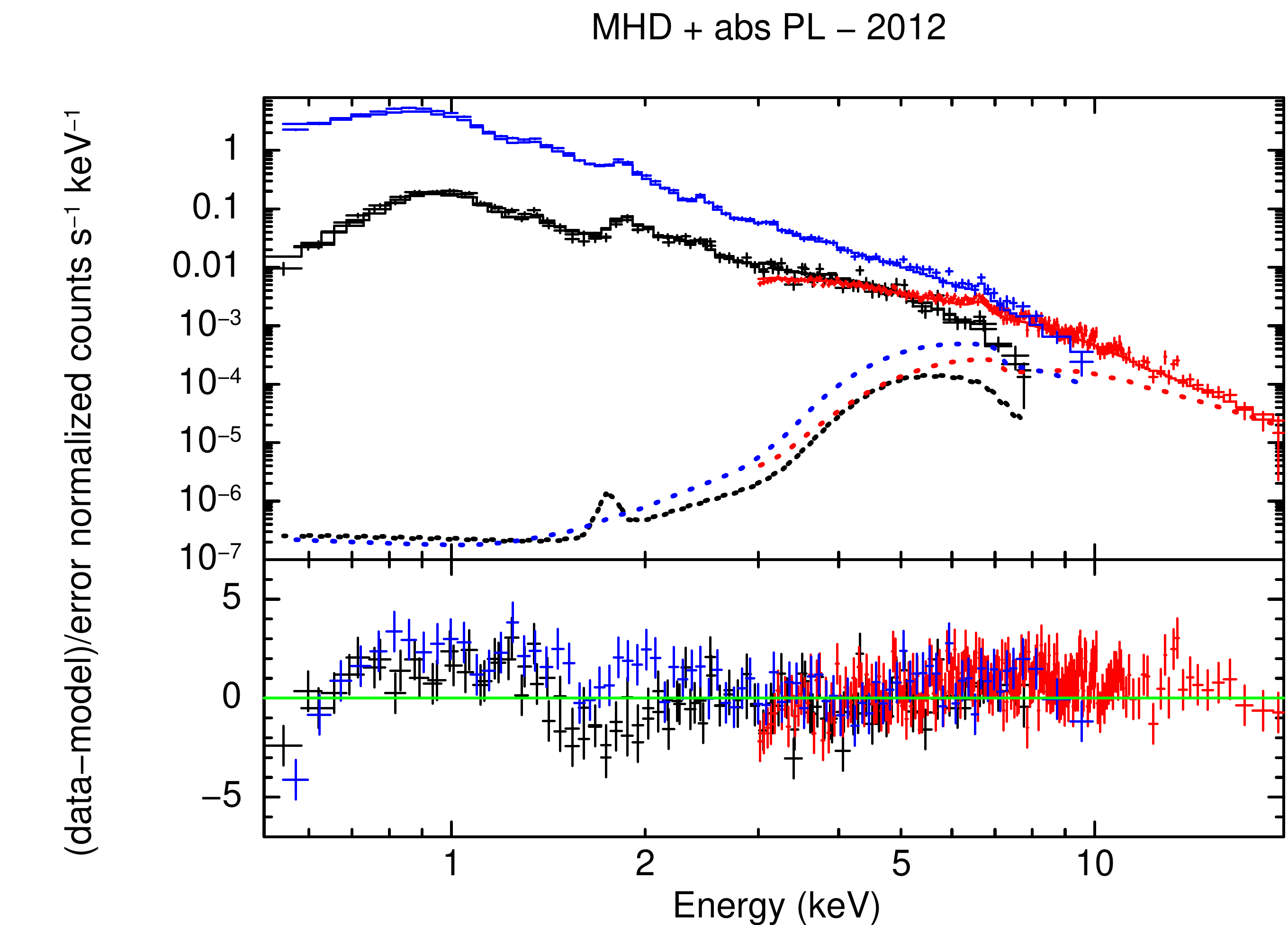}
    \end{minipage}
    \hfill
    \begin{minipage}{0.32\textwidth}
    \includegraphics[angle=270,width=\textwidth]{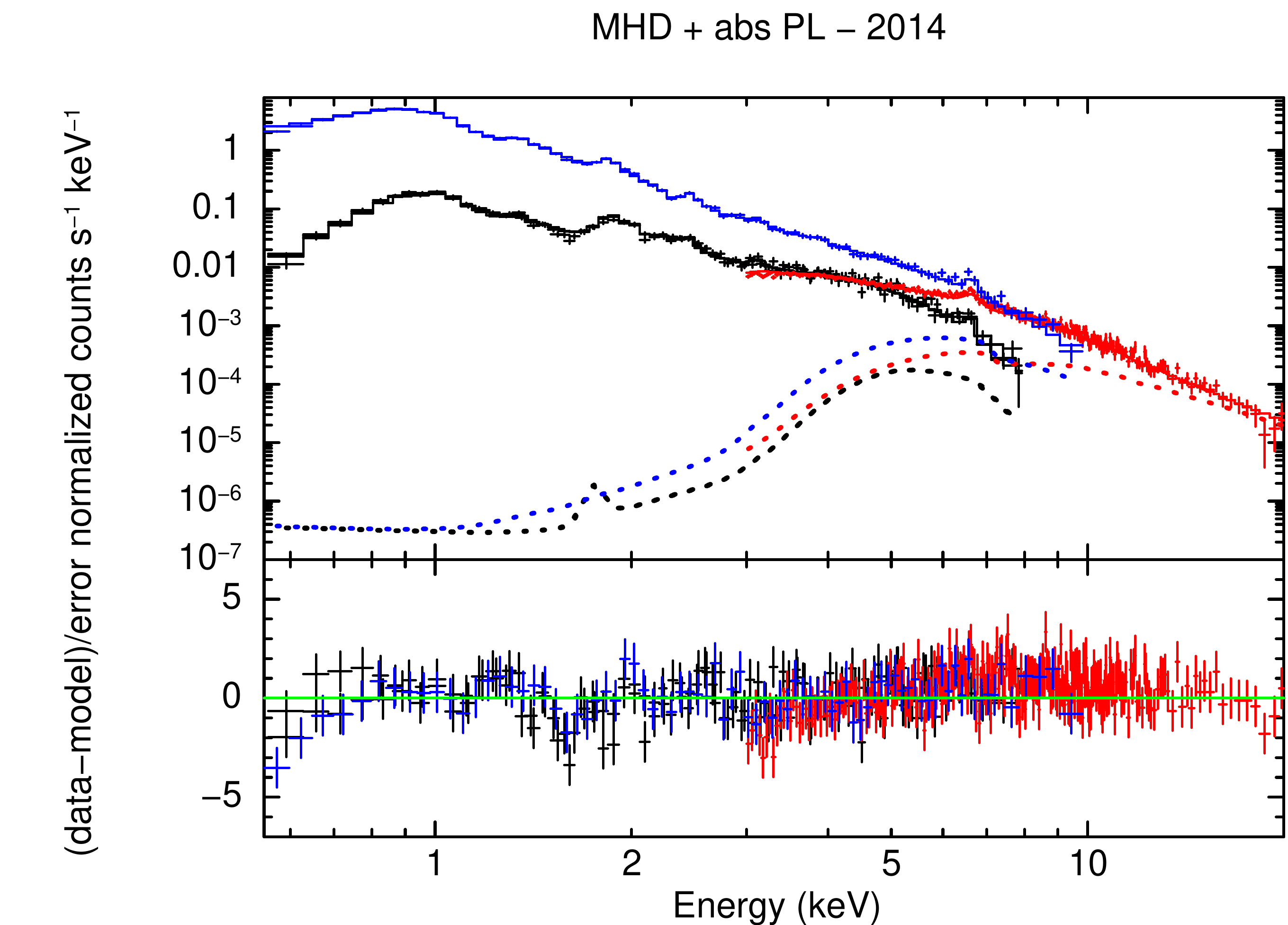}
    \end{minipage}
    \hfill
    \begin{minipage}{0.32\textwidth}
    \includegraphics[angle=270,width=\textwidth]{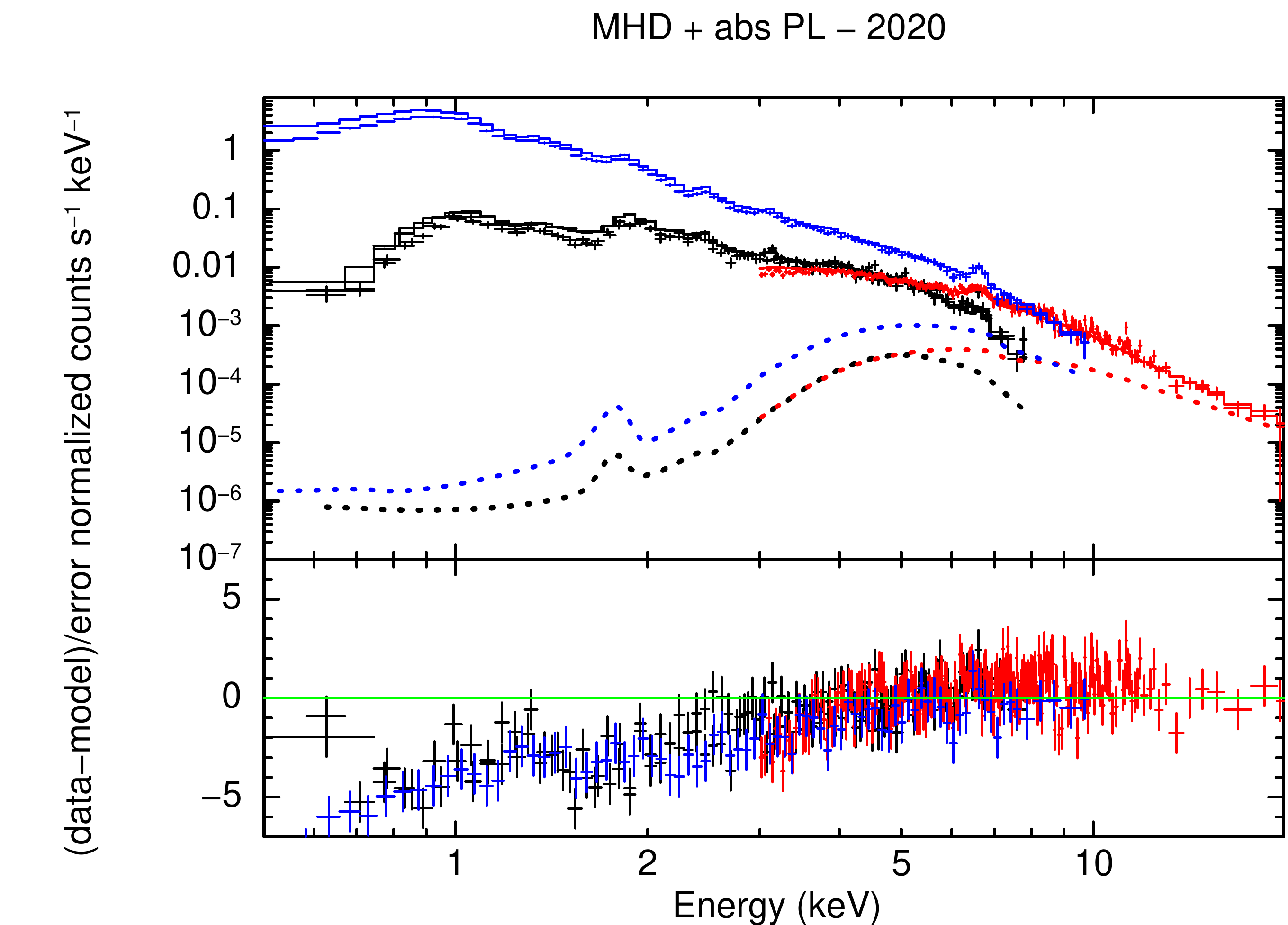}
    \end{minipage}
    
    \caption{Same as Fig. \ref{fig:fenom} but \xmm/RGS1 spectra are shown separately in the first row. Black, blue and red points are \chandra/ACIS-S, \xmm/pn and \nustar/FPMA,B spectra, respectively. \emph{First and second row.} The spectra are compared with the MHD model by Or20 (solid lines). \emph{Last row.} The spectra are compared with the MHD model by Or20 plus the PWN component found in this paper (solid lines). The contribution of the absorbed PL folded with each response matrix is also showed with dashed lines with the same color of the corresponding data points.}
    \label{fig:synth}
\end{figure*}

We did not perform any fit on the model parameters\footnote{We only associate a systematic error of $5\%$ on the synthetic spectra when comparing them to the \xmm/pn data because of the very small error bars of data set collected by this detector.}, but we just compared the spectra as they are synthesized from model B18.3 with the actual ones at all epochs and without any renormalization of the model or ad-hoc assumptions. In comparing synthetic and actual spectra, we only left the factors $C_i$ free to vary, to account for cross-calibration issues. As done in the standard analysis, we put $C_{pn}=1$ (for the \xmm/pn spectra) and all other $C_i$ were forced to stay within their typical ranges (\citealt{mhm15}). The comparison between the thermal emission predicted by model B18.3 and the actual data for each epoch is shown in the first row of Fig. \ref{fig:synth} for the \xmm/RGS1 (1st order) and in the second row for \xmm/pn, \chandra/ACIS-S and \nustar/FPMA. 

We stress that the plots in Fig. \ref{fig:synth} were obtained without performing any fit, except for the cross-calibration constants. It is striking that the model is able to well reproduce the X-ray emission below 10 keV at all epochs and for all instruments, even the main lines and the continuum emission observed by RGS1. This strongly suggests that model B18.3 accurately describes the thermal contribution to X-rays from the shock-heated plasma and its evolution in the time lapse considered. 

The good description of the line widths that we see in the first row of Fig. \ref{fig:synth} also indicates that the spatial distribution of macroscopic plasma velocity (mainly responsible for the line broadening) is also well reproduced by model B18.3, further confirming the pertinence of this model. Model B18.3 only slightly overestimates the flux in the 2020 data, as also shown by the X-ray light curves in Or20.

Furthermore, Fig. \ref{fig:synth} shows that model B18.3 self-consistently describes the rapid evolution of thermal emission in the $0.5-8$ keV band in the $2012-2020$ time range. In particular, it accounts for the increase in the emission measure of the hot plasma associated with the expansion of the shock in the HII region and with the hardening of X-ray spectra reported by \citet{fzp16}, S21 and \citet{rpz21}. However, we also note that the thermal emission synthesized from model B18.3 underestimates the X-ray flux above $\sim 8$ keV. Interestingly, the difference between synthetic thermal emission and actual X-ray spectra above $8$ keV does not evolve, but stays constant between 2012 and 2014. This indicates that an additional steady, hard X-ray emitting component needs to be included. 

Since this model already takes into account all possible contributions from thermal shock-heated plasma and self-consistently describes their temporal evolution, in the light of the results of Sect.~\ref{sub:spectra_analysis}, it is natural to assert that the radiation in the hard energy part of the X-ray spectra most likely includes a steady non-thermal component.
This strongly corroborates the result of our spectral analysis, which showed that a non-thermal component is needed to achieve the best description of the data.

As a further step, we compared the actual spectra with the thermal emission synthesized from model B18.3 plus the absorbed PL component inferred from the spectral analysis in the previous section (and consistent with that found by G21), representing the emission from the PWN. To this end, we assumed the same PL parameters as those found in the simultaneous multi-epoch fit, namely those that are best constrained: $\Gamma = 2.8$ and norm =$7.1 \times 10^{-4}$ ph/s/keV/cm$^{2}$. We verified that, by changing the norm value in the range [4-8]$\times 10^{-4}$ ph/s/keV/cm$^{2}$ the results do not significantly change. Coherently with the standard analysis, the foreground absorption component is kept fixed to 2.35 $\times 10^{21}$ cm$^{-2}$ (\citealt{pzb06}). Again, we did not perform any fit to fine tune the photon index and the normalization of the PL. The third row in Fig. \ref{fig:synth} compares the synthetic spectra derived from model B18.3 plus the absorbed PL with the actual spectra at the three epochs. In the bottom rows of Table \ref{tab:fit_whole}, we report the values of $\chi^2$ obtained when comparing the actual data with the emission synthesized either from model B18.3 alone or from model B18.3 plus the absorbed PL. 

Analogously to what we found with the standard analysis, an additional component described by a steady and heavily absorbed PL is needed for a best description of the actual data. This is a further indication of the non-thermal nature of the dominant emitting component above 10 keV and of heavy absorption of this component compatible with being embedded in the innermost ejecta.

\subsection{A deeper look into the putative PWN} \label{sect:putPWN}
In this section we made use of available spectral information to give a closer look to the elusive PWN and pulsar possibly hidden inside \sna, with the main purpose of determining a range for the pulsar spin-down luminosity, $\dot{E}$, and period, $P$, to be compared with other known objects.

We built a simple model for the PWN synchrotron spectrum considering, as it is generally done, a broken power-law with energy flux given by  $S_\nu(\nu)\propto \nu^{-\alpha_1}$ for $\nu \leq \nu_B$ and $S_\nu(\nu)\propto \nu^{-\alpha_2}$ otherwise, with a main spectral break at a frequency $\nu_B$ \citep [see e.g.][for a review on PWNe properties]{Gaensler:2006,Slane:2017}.
PWNe have been shown to be characterized by a rather flat radio spectrum at all frequencies below $\nu_B$, with an energy spectral index in the range $\alpha_1\simeq 0-0.4$ \citep [see e.g.][]{Reynolds:2017}.
On the other hand the emission at higher energies,  is well described by a steeper PL, with a photon index in the X-ray band of $\Gamma_2\approx 2.2$ (with $\Gamma=1+\alpha$).
Apart from the spectral break separating the two populations of particles responsible for the low and high energy emissions, a number of other spectral variations are observed from optical to gamma-ray bands, with some evidence for a spectral steepening with increasing distance from the PWN center at the X-rays \citep [see ][and references therein]{ Reynolds:2017}.

For the present work we have assumed $\alpha_1=0.3$ for the low energy spectral component. 
However we verified that considering a lower value causes a negligible variation to the final estimate of the pulsar spin-down luminosity and period.
The analysis of the X-ray spectra, as discussed above in this section (see Sect.~\ref{sub:spectra_analysis} and Table \ref{tab:fit_whole} therein), gave us an indication of the non-thermal photon index in the hard X-ray band ($\Gamma=2.8$ in 10-20 keV). %
Despite being slightly higher than the  one usually found for X-ray spectra of PWNe, it is still compatible with previous observations. This for the following main points:   i) the PWN of \sna\ would be the youngest ever found, likely characterized by higher magnetic field than other objects \citep{Torres:2014}; ii) the typical X-ray $\Gamma$ is usually measured in the 0.5-8 keV band, while here we are able to measure only the very hard part of the spectrum, whose slope might be modified by the vicinity of the synchrotron cut-off.
The frequency of the spectral break is indeed left as a free parameter of the model.

The only other available observational information on the properties of the putative PWN comes from ALMA's measurements of a radio blob (of radius $0.01$ pc), detected at 679 GHz with a luminosity in the range $L_R=40\ls-90 \ls$, and located at the position where the pulsar supposedly formed \citep{cmg19}.
Actually this radio emission is very difficult to be interpreted as a clear evidence for the PWN, since it might even be of a completely different nature, as already pointed out by  \citet{cmg19}.
In any case the range of radio luminosities detected by ALMA can be used as a reliable upper limit for the PWN luminosity at 679 GHz, since a possible higher emission at this specific frequency should have been detected as well.

By moving the break frequency in a wide range $10^{13}-10^{16}$ Hz  and considering the limits imposed by Cigan's radio luminosities and available X-ray data, we obtain a range for the bolometric luminosity of the synchrotron spectrum of $\sim[0.15,\,1.5]\times10^{37}$ erg/s.

This must then be converted into a range for the pulsar spin-down power. In order to do that, one has to assume a value for the conversion efficiency from rotational power into radiation ($\eta$). Unfortunately this parameter is poorly constrained, with some evidence for the Crab nebula -- believed to be one of the most efficient sources -- to be of $\eta\sim30\%$ \citep{h2008}.
Here, in order to be as much general as possible, we have considered a minimum conversion efficiency of $\eta=5\%$ (which is very unlikely, meaning that almost all the energy is lost somewhere else) and a maximum one of $\eta=70\%$. This translates in a wide range of possible pulsar spin-down luminosities:
$\dot{E}\in[0.3\,,30]\times10^{37}$erg/s.
\begin{figure}
    \centering
    \includegraphics[width=.96\columnwidth]{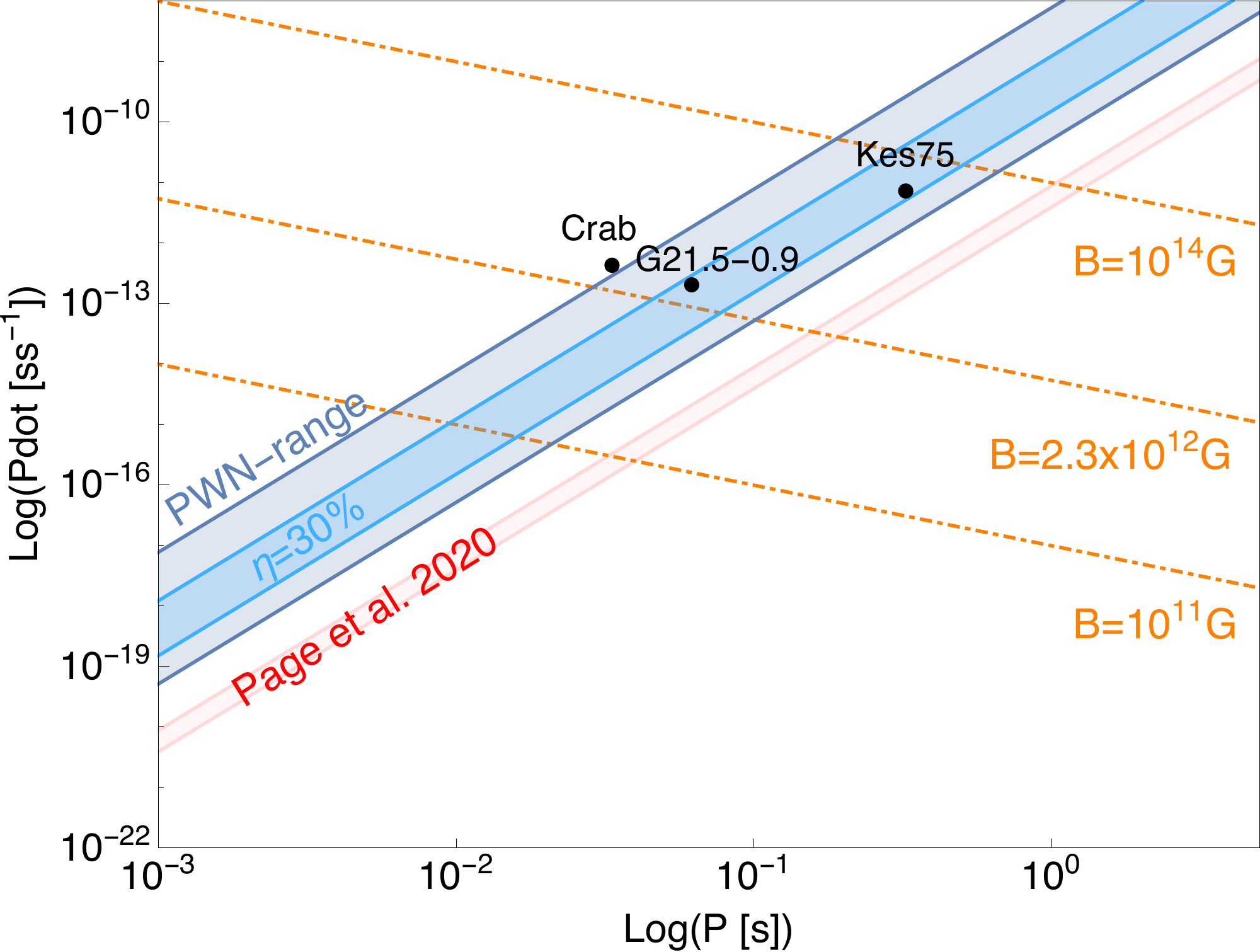}
    \caption{Position of the estimated range of possible spin-down luminosity of the putative pulsar in the $P-\dot{P}$ diagram (in blue, in light blue the one corresponding to a conversion efficiency $\eta=30\%$). 
    Orange dot-dashed lines indicate a large range for magnetic fields associated with pulsars, plus the value computed as average of the population of pulsars associated with X-ray PWNs \citep{K&P2008}  (namely $2.3\times10^{12}$G).
    The range for the pulsar position as found in \citet{pbg20} is also reported for comparison as a red shaded area.
    The position of three young pulsars associated with young PWNe (namely the Crab Nebula, G21.5-0.9 and Kes75) is shown with black dots.}
    \label{fig:PPdot}
\end{figure}
In Fig.~\ref{fig:PPdot} we show how this range can be converted in a region of the pulsars $P-\dot{P}$ diagram (the blue shaded area), making use of the relation: $\dot{E}=4\pi^2 I \dot{P}/P^3$, where $I\sim 10^{45}$ g$\,$cm$^{2}$ is the pulsar momentum of inertia and $\dot{P}$ the period time derivative. 
The region corresponding to a conversion efficiency of $\eta=30\%$ is also shown (light blue). 
For a better comparison we report the position of the only three young systems known at present (with $t_{\rm{age}}\lesssim 1000$ yr), namely: the Crab Nebula, G21.5-0.9 and Kes75 \citep{h2008,gv2000,bb2008}. 
Older sources cannot be consistently used to compare with this putative pulsar, since their long term evolution changed strongly their birth properties. 

As it can be easily noticed, the range we found for the pulsar position in the $P$-$\dot{P}$ diagram is perfectly consistent with that of known pulsars powering young systems.
Moreover it must be noticed that the apparent inconsistency with the range estimated in  \citet{pbg20} (shown as the red shaded area in Fig.~\ref{fig:PPdot}), is due to the fact that the authors used the range of luminosities measured at 679 GHz ($L_R$) to estimate the possible positioning of the pulsar in the P-Pdot diagram (see \citealt{pbg20}, Fig. 1), not considering that only a fraction of the spin-down power is converted into radiation \citep[see e.g.][]{bbcrab2014}. 
Their range must be then considered as a lower limit for the spin-down luminosity: if the blob is illuminated by the pulsar -- or if it is directly the evidence for the emerging nebula -- it is clear that the spin-down luminosity cannot be lower than the observed $L_R$.

Fig.~\ref{fig:PPdot} shows as well the lines corresponding to a wide range of magnetic fields characteristic of the pulsar population ($10^{11}-10^{14}$ G), using the relation $B=3.2 \times 10^{19} (P\dot{P})^{1/2}$ G. The intermediate line shows the average magnetic field as computed from the known population of X-ray emitting PWNe ($B\simeq 2.3\times 10^{12}$ G, from \citealt{K&P2008}).
If the putative pulsar is characterized by a magnetic field close to the average value of pulsars associated with X-ray emitting PWNe,  its period is expected in the range $P\in[30,100]$ ms.
Having a conversion efficiency not so lower than that usually assumed for the Crab ($30\%$), this range reduces to $P\in[45,80]$ ms.

\section{Discussion} \label{sect:disc}
In this work, we demonstrated that, at all epochs analyzed, the description of the broadband X-ray emission of \sna\  requires a PL non-thermal component with absorption compatible with that of the cold and dense metal-rich ejecta in the innermost part of \sna.
The non-thermal nature of this high energy source is supported by the temporal evolution of the 10-20 keV flux. In fact, the flux evolution of \sna\ has been dramatically changing in the last 2 decades on a time scale of a few years or even months (see Fig. 4 in A21 and references therein). In particular, the soft X-ray emission is currently fading away with time, while the hard one is still increasing, even if at a slower pace. Both these trends are in stark contrast with the 10\% flux variation in the 10-20 keV band. It is hard to explain how the thermal emission at all other bands significantly changed year by year, while it remained almost constant in the 10-20 keV band. On the other hand, the emission from a PWN is expected not to vary significantly on such a short time scale (much smaller than the pulsar spin down age $\tau_0\sim$ kyr).

The measured photon index $\Gamma \sim 2.8$ is compatible with the synchrotron emission from a PWN.
DSA working at the outer shell of the remnant could also be invoked as a possible source of this radiation (see G21). We found that the addition of an unabsorbed PL to the \emph{3-kT model} in fact improves the fit quality of the multi-epoch observations.
However, the best description of the spectra is achieved when heavy absorption of the PL is included, thus suggesting that the source of the non-thermal emission is located close to the center of the remnant, instead of the the border as needed in the DSA case.
Furthermore, DSA requires a similar temporal evolution for radio and X-ray fluxes, while  the flux in the 10-20 keV band has slightly ($\sim 10$\%) increased between 2014 and 2020, and it is consistent with being constant at the 90\% confidence level. 
On the contrary, the flux in the radio band has increased by $\sim 60\%$ (and, in general, has dramatically changed at all other bands) in the same time lapse (see Fig. 4 in A21 and \citealt{cgn18}). 
In the light of these considerations we can exclude standard DSA as the mechanism producing the observed radiation, that we thus interpret as the result of non-thermal emission from a PWN located at the heart of \sna.

\begin{figure*}[!ht]
    \includegraphics[width=.95\textwidth]{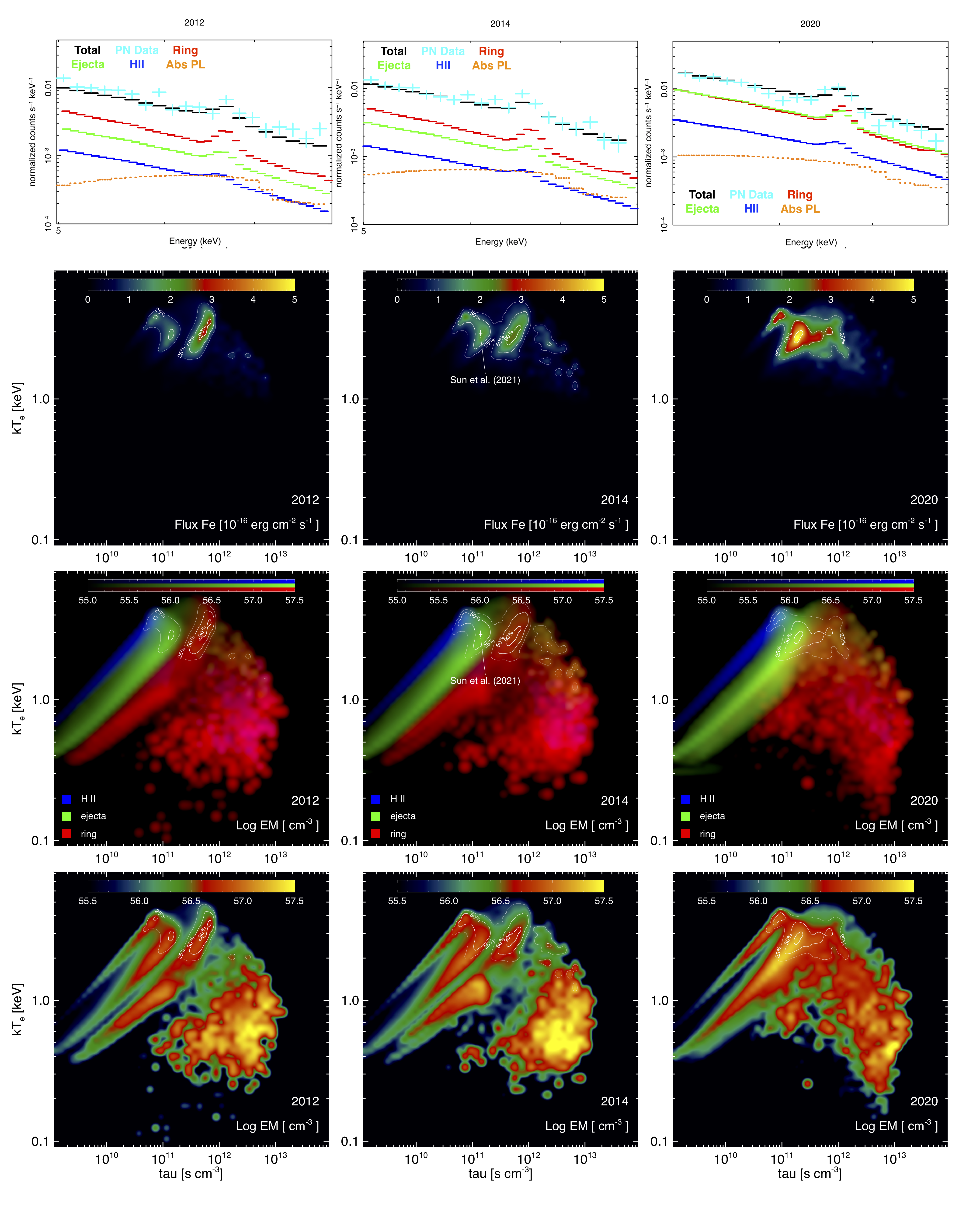}

    \caption{Comparison between Fe K observed emission lines and the various contributions estimated from the B18.3 model. \emph{From top to bottom}. First panel shows the synthetic spectra of the ring (red), ejecta (green), HII region (blue), absorbed PL (orange), the total synthetic spectrum (black) and the actual \xmm/pn data (cyan); the second panel shows the continuum-subtracted Fe K line flux maps; the third and fourth panels show the distribution of emission measure maps for \sna\ in logarithmic units, isolating the various components of the plasma or showing the total distribution, respectively. \emph{From left to right.} First column pertains to 2012, the second to 2014, the third to 2020. White contours identify the regions where the 25\%, 50\% and 90\% of Fe K is originated. The panels relative to 2014 also provide a comparison between the maps and the findings by S21 for the three-thermal model.}
    \label{fig:FeK}
\end{figure*}

We investigated the nature of the $\sim10$\% variation observed in the best-fit 10-20 keV flux making use of high resolution MHD simulations of \sna. In particular we exploited the MHD model B18.3 from Or20 and evaluated, in the time lapse analyzed, the variation of thermal flux arising from the shock-heated plasma in the 10-20 keV band and the decrease of the absorbing power by the innermost cold ejecta due to the remnant expansion. We found that both these effects concur, with similar weights, to the $\sim 10\%$ increase of the flux in the 10-20 keV band, thus fitting nicely in the scenario of a PWN embedded by cold ejecta in the heart of \sna. 
In particular, we found that the absorbed flux of the PL in correspondence with the Fe K emission line increases by a factor $\sim$ 5 between 2014 and 2020 (see first row in Fig. \ref{fig:FeK}), a clear signature of the decrease of the absorption of the expanding innermost ejecta. 

To investigate also the possible thermal (blackbody) contribution of the NS embedded in the inner part of \sna, we replaced the PL component with a blackbody (\texttt{bbodyrad} model in XSPEC). We found no improvements in the fit quality with respect to the 3-kT model for typical radius ($2 < R < 20 $ km) of the NS or CCO (\citealt{dlu08}). Conversely, we found a significant improvement in the $\chi^2$ for temperature kT$_{\rm{bb}} \sim 4$ keV and with a $R$ of a few meters. However, the spectral shape of such absorbed blackbody is practically indistinguishable from the absorbed PL, being all the radiation below $\sim 8$ keV suppressed by absorption from cold and dense ejecta. 
Given the unrealistically small radius found for the best-fit blackbody, we conclude that our X-ray spectra do not contain any hint for the NS radiation. Detailed analysis on the time evolution of the thermal X-ray emission from the NS will be discussed in a companion paper (Dohi et al., in preparation).

A21 stated that including a PL component leads to an unreasonably high abundance of Fe to reproduce the Fe K emission line. From the best-fit parameters in Table \ref{tab:fit_whole}, it can be seen that by considering a PL emission absorbed by the innermost ejecta, the Fe abundance is fully consistent with the typical values found for the LMC, and with previous estimates (S21, A21). To further investigate the different plasma components contributing to the Fe K line, we produced synthetic \xmm/pn spectra from model B18.3 separating the contributions from the three main plasma components of \sna: the shocked material from the ring, from the HII region and the shocked ejecta (see Fig. \ref{fig:FeK}). We then included also the absorbed PL and summed all the contributions to generate the synthetic total spectrum (black data points in the first row of Fig. \ref{fig:FeK}). Finally, we compared the resulting spectrum with the observed one at each epoch (first row of Fig. \ref{fig:FeK}).

The observed Fe K flux can be correctly recovered by including the non-thermal component. The synthetic spectra in Fig. \ref{fig:FeK} show how the different X-ray emitting regions of \sna\ contribute to the Fe emission at different epochs. While in 2012 the Fe K emission is mainly due to the shocked ring, especially to the lower density material lying between the dense clumps, in 2020 the ejecta contribution to the emission line is as high as that of the ring. To better highlight this trend, we produced continuum-subtracted line flux maps of the Fe K line from model B18.3, and compared it to the total EM distribution (second, third and fourth rows in Fig. \ref{fig:FeK}, respectively). The contours traced help to identify the origin of the Fe K flux. The contribution to Fe K flux from the shocked ejecta and shocked material from the HII region gradually increases from 2012 to 2020; this contribution is characterized by an average value of the ionization parameter significantly lower than that of the contribution due to the shocked material from the ring.
For reference, a comparison with the hard thermal component reported by S21 for 2014 data is also shown. In the light of comparison between the MHD model and the observations, we suggest that, in the next few years, the main contribution to the Fe K line will come from the shocked ejecta lying in the outer envelope of \sna. 

\section{Summary and conclusions}
\label{sec:sum}

In this paper we showed that the most likely and physically reasonable origin of the hard X-ray emission ($> 10$~keV) observed with \nustar\ is non-thermal radiation emitted by an heavily absorbed PWN embedded in the cold and dense ejecta of \sna. Our findings can be summarized a follows.

\begin{enumerate}
    \item The standard X-ray analysis of \xmm, \chandra\ and \nustar\ observations based on a purely thermal model shows significant residuals in the [$10-20$]~keV band at all epochs. By including a power-law, absorbed by the cold inner ejecta, these residuals disappear.
    \item The best-fit non-thermal component is characterized by a photon index $\Gamma \sim 2.8$, which is compatible with being synchrotron emission. The spectral parameters of the absorbed power-law are compatible with being constant at all epochs, indicating that this component is almost constant over a time-lapse of eight years (2012-2020). An unabsorbed power-law, representing the DSA scenario, provides worse description of the spectra, indicating that the source of the non-thermal emission is embedded in the hearth of SN 1987A. All the above lines of evidence favor a PWN rather than DSA as the origin of the flux excess above 10~keV. 
    \item The comparison of the observed spectra with those synthesized from a state-of-the-art MHD model of \sna\ (Or20) shows again a significant flux excess in [$10-20$]~keV band. The MHD model describe only the thermal contribution to the emission. By adding the absorbed PL component, derived independently with the standard data analysis, to the synthetic spectra from the MHD model, the resulting spectra are compatible with the observed broad-band X-ray spectra of \sna\ at all epochs analyzed. 
    \item The 10-20 keV flux slightly increases by $\sim 10\%$ in the time-lapse analyzed. By exploiting the MHD model of \sna, we found that this increase is entirely ascribable, with similar percentages, to the decrease of absorption by the cold and dense ejecta and to the increase of the thermal contamination in the high-energy band. In the same time-lapse, the radio emission due to DSA in the outer shell of the remnant is observed to increase by $\sim$60\% (\citealt{cgn18}). These lines of evidence exclude DSA as a possible mechanism powering the observed non-thermal emission.
    \item The properties found for the X-ray non-thermal component and the upper limit on the radio luminosity of a putative PWN derived from ALMA's observations are fully compatible with a PWN spectrum described by a standard broken PL.
    \item We derived the ranges of spin-down luminosity and period of the putative pulsar, associated with the modeled synchrotron spectrum. We found values that are fully compatible with those inferred for other known young PWNe (i.e., the Crab Nebula, G21.5-0.9 and Kes75), thus leaving large room for the existence of a PWN inside \sna.
\end{enumerate}

\software{CIAO \citep{ciao}, HEASOFT\footnote{https://heasarc.gsfc.nasa.gov/docs/software/heasoft/}, XSPEC \citep{arn96}, NuSTARDAS\footnote{https://heasarc.gsfc.nasa.gov/docs/nustar/analysis/nustar\_swguide.pdf}, SAS \citep{sas}, DS9\footnote{https://sites.google.com/cfa.harvard.edu/saoimageds9/home?authuser=0}, PLUTO\footnote{http://plutocode.ph.unito.it} \citep{2012ApJS..198....7M}.  }

\facilities{\chandra\ \footnote{https://cxc.harvard.edu/index.html}, \nustar\ \citep{NuSTAR}, \xmm\ \citep{jla01}}

\section*{Acknowledgements}
This research has made use of the NuSTAR Data Analysis Software (NuSTARDAS) jointly developed by the ASI Space Science Data Center (SSDC, Italy) and the California Institute of Technology (Caltech, USA). This project has received funding from the European Union’s Horizon 2020 research and innovation program under grant agreement No. 101004131 (SHARP). The MHD model was implemented with the PLUTO code developed at the Turin Astronomical Observatory (Italy) in collaboration with the Department of General Physics of Turin University (Italy) and the SCAI Department of CINECA (Italy). We acknowledge that the results of this research have been achieved using the PRACE Research Infrastructure resource Marconi based in Italy at CINECA (PRACE Award N.2016153460). Additional computations were carried out at the SCAN (Sistema di Calcolo per l'Astrofisica Numerica) facility for high performance computing at INAF-Osservatorio Astronomico di Palermo. EG, MM, SO, BO, FB and GP acknowledge financial contribution from the PRIN INAF 2019 grant "From massive stars to supernovae and supernova remnants: driving mass, energy and cosmic rays in our Galaxy" and the INAF mainstream program "Understanding particle acceleration in galactic sources in the CTA era". This work is supported by JSPS Grants-in-Aid for Scientic Research "KAKENHI" (A: Grant Number JP19H00693; B: Grant Number JP21K03545). SN and MO acknowledge supports from Pioneering Program of RIKEN for Evolution of Matter in the Universe (r-EMU) and RIKEN Interdisciplinary Theoretical and Mathematical Sciences Program (iTHEMS). SN also acknowledges the support from Pioneering Program of RIKEN for Evolution of Matter in the Universe (r-EMU).
BO acknowledges Elena Amato, Niccolò Bucciantini and Rino Bandiera for fruitful discussion.

\bibliography{references}

\begin{thebibliography}{}
\expandafter\ifx\csname natexlab\endcsname\relax\def\natexlab#1{#1}\fi
\providecommand{\url}[1]{\href{#1}{#1}}
\providecommand{\dodoi}[1]{doi:~\href{http://doi.org/#1}{\nolinkurl{#1}}}
\providecommand{\doeprint}[1]{\href{http://ascl.net/#1}{\nolinkurl{http://ascl.net/#1}}}
\providecommand{\doarXiv}[1]{\href{https://arxiv.org/abs/#1}{\nolinkurl{https://arxiv.org/abs/#1}}}

\bibitem[{{Alp} {et~al.}(2021){Alp}, {Larsson}, \& {Fransson}}]{alf21}
{Alp}, D., {Larsson}, J., \& {Fransson}, C. 2021, arXiv e-prints,
  arXiv:2103.02612.
\newblock \doarXiv{2103.02612}

\bibitem[{{Alp} {et~al.}(2018){Alp}, {Larsson}, {Fransson}, {Indebetouw},
  {Jerkstrand}, {Ahola}, {Burrows}, {Challis}, {Cigan}, {Cikota}, {Kirshner},
  {van Loon}, {Mattila}, {Ng}, {Park}, {Spyromilio}, {Woosley}, {Baes},
  {Bouchet}, {Chevalier}, {Frank}, {Gaensler}, {Gomez}, {Janka}, {Leibundgut},
  {Lundqvist}, {Marcaide}, {Matsuura}, {Sollerman}, {Sonneborn},
  {Staveley-Smith}, {Zanardo}, {Gabler}, {Taddia}, \& {Wheeler}}]{alf18a}
{Alp}, D., {Larsson}, J., {Fransson}, C., {et~al.} 2018, \apj, 864, 174,
  \dodoi{10.3847/1538-4357/aad739}

\bibitem[{{Arnaud}(1996)}]{arn96}
{Arnaud}, K.~A. 1996, in ASP Conf. Ser. 101: Astronomical Data Analysis
  Software and Systems V, 17

\bibitem[{{Bietenholz} \& {Bartel}(2008)}]{bb2008}
{Bietenholz}, M.~F., \& {Bartel}, N. 2008, \mnras, 386, 1411,
  \dodoi{10.1111/j.1365-2966.2008.13058.x}

\bibitem[{{Bionta} {et~al.}(1987){Bionta}, {Blewitt}, {Bratton}, {Casper},
  {Ciocio}, {Claus}, {Cortez}, {Crouch}, {Dye}, {Errede}, {Foster}, {Gajewski},
  {Ganezer}, {Goldhaber}, {Haines}, {Jones}, {Kielczewska}, {Kropp}, {Learned},
  {Losecco}, {Matthews}, {Miller}, {Mudan}, {Park}, {Price}, {Reines},
  {Schultz}, {Seidel}, {Shumard}, {Sinclair}, {Sobel}, {Stone}, {Sulak},
  {Svoboda}, {Thornton}, {van der Velde}, \& {Wuest}}]{bbb87}
{Bionta}, R.~M., {Blewitt}, G., {Bratton}, C.~B., {et~al.} 1987, \prl, 58,
  1494, \dodoi{10.1103/PhysRevLett.58.1494}

\bibitem[{{Borkowski} {et~al.}(1997){Borkowski}, {Blondin}, \&
  {McCray}}]{bbm97}
{Borkowski}, K.~J., {Blondin}, J.~M., \& {McCray}, R. 1997, \apj, 477, 281,
  \dodoi{10.1086/303691}

\bibitem[{{B{\"u}hler} \& {Blandford}(2014)}]{bbcrab2014}
{B{\"u}hler}, R., \& {Blandford}, R. 2014, Reports on Progress in Physics, 77,
  066901, \dodoi{10.1088/0034-4885/77/6/066901}

\bibitem[{{Cendes} {et~al.}(2018){Cendes}, {Gaensler}, {Ng}, {Zanardo},
  {Staveley-Smith}, \& {Tzioumis}}]{cgn18}
{Cendes}, Y., {Gaensler}, B.~M., {Ng}, C.~Y., {et~al.} 2018, \apj, 867, 65,
  \dodoi{10.3847/1538-4357/aae261}

\bibitem[{{Cigan} {et~al.}(2019){Cigan}, {Matsuura}, {Gomez}, {Indebetouw},
  {Abell{\'a}n}, {Gabler}, {Richards}, {Alp}, {Davis}, {Janka}, {Spyromilio},
  {Barlow}, {Burrows}, {Dwek}, {Fransson}, {Gaensler}, {Larsson}, {Bouchet},
  {Lundqvist}, {Marcaide}, {Ng}, {Park}, {Roche}, {van Loon}, {Wheeler}, \&
  {Zanardo}}]{cmg19}
{Cigan}, P., {Matsuura}, M., {Gomez}, H.~L., {et~al.} 2019, \apj, 886, 51,
  \dodoi{10.3847/1538-4357/ab4b46}

\bibitem[{{de Luca}(2008)}]{dlu08}
{de Luca}, A. 2008, in American Institute of Physics Conference Series, Vol.
  983, 40 Years of Pulsars: Millisecond Pulsars, Magnetars and More, ed.
  C.~{Bassa}, Z.~{Wang}, A.~{Cumming}, \& V.~M. {Kaspi}, 311--319,
  \dodoi{10.1063/1.2900173}

\bibitem[{{Esposito} {et~al.}(2018){Esposito}, {Rea}, {Lazzati}, {Matsuura},
  {Perna}, \& {Pons}}]{erl18}
{Esposito}, P., {Rea}, N., {Lazzati}, D., {et~al.} 2018, \apj, 857, 58,
  \dodoi{10.3847/1538-4357/aab6b6}

\bibitem[{{Frank} {et~al.}(2016){Frank}, {Zhekov}, {Park}, {McCray}, {Dwek}, \&
  {Burrows}}]{fzp16}
{Frank}, K.~A., {Zhekov}, S.~A., {Park}, S., {et~al.} 2016, \apj, 829, 40,
  \dodoi{10.3847/0004-637X/829/1/40}

\bibitem[{{Fransson} \& {Chevalier}(1987)}]{fcr87}
{Fransson}, C., \& {Chevalier}, R.~A. 1987, \apjl, 322, L15,
  \dodoi{10.1086/185028}

\bibitem[{{Fruscione} {et~al.}(2006){Fruscione}, {McDowell}, {Allen},
  {Brickhouse}, {Burke}, {Davis}, {Durham}, {Elvis}, {Galle}, {Harris},
  {Huenemoerder}, {Houck}, {Ishibashi}, {Karovska}, {Nicastro}, {Noble},
  {Nowak}, {Primini}, {Siemiginowska}, {Smith}, \& {Wise}}]{ciao}
{Fruscione}, A., {McDowell}, J.~C., {Allen}, G.~E., {et~al.} 2006, in Society
  of Photo-Optical Instrumentation Engineers (SPIE) Conference Series, Vol.
  6270, Society of Photo-Optical Instrumentation Engineers (SPIE) Conference
  Series, ed. D.~R. {Silva} \& R.~E. {Doxsey}, 62701V,
  \dodoi{10.1117/12.671760}

\bibitem[{{Gabriel} {et~al.}(2004){Gabriel}, {Denby}, {Fyfe}, {Hoar}, {Ibarra},
  {Ojero}, {Osborne}, {Saxton}, {Lammers}, \& {Vacanti}}]{sas}
{Gabriel}, C., {Denby}, M., {Fyfe}, D.~J., {et~al.} 2004, in Astronomical
  Society of the Pacific Conference Series, Vol. 314, Astronomical Data
  Analysis Software and Systems (ADASS) XIII, ed. F.~{Ochsenbein}, M.~G.
  {Allen}, \& D.~{Egret}, 759

\bibitem[{{Gaensler} \& {Slane}(2006)}]{Gaensler:2006}
{Gaensler}, B.~M., \& {Slane}, P.~O. 2006, \araa, 44, 17,
  \dodoi{10.1146/annurev.astro.44.051905.092528PDF:
  http://arjournals.annualreviews.org/doi/pdf/10.1146/annurev.astro.44.051905.092528}

\bibitem[{{Gotthelf} {et~al.}(2000){Gotthelf}, {Vasisht}, {Boylan-Kolchin}, \&
  {Torii}}]{gv2000}
{Gotthelf}, E.~V., {Vasisht}, G., {Boylan-Kolchin}, M., \& {Torii}, K. 2000,
  \apjl, 542, L37, \dodoi{10.1086/312923}

\bibitem[{{Greco} {et~al.}(2021){Greco}, {Miceli}, {Orlando}, {Olmi},
  {Bocchino}, {Nagataki}, {Ono}, {Dohi}, \& {Peres}}]{gmo21}
{Greco}, E., {Miceli}, M., {Orlando}, S., {et~al.} 2021, \apjl, 908, L45,
  \dodoi{10.3847/2041-8213/abdf5a}

\bibitem[{{Haberl} {et~al.}(2006){Haberl}, {Geppert}, {Aschenbach}, \&
  {Hasinger}}]{hga06}
{Haberl}, F., {Geppert}, U., {Aschenbach}, B., \& {Hasinger}, G. 2006, \aap,
  460, 811, \dodoi{10.1051/0004-6361:20066198}

\bibitem[{{Harrison} {et~al.}(2013){Harrison}, {Craig}, {Christensen},
  {Hailey}, {Zhang}, {Boggs}, {Stern}, {Cook}, {Forster}, {Giommi},
  {Grefenstette}, {Kim}, {Kitaguchi}, {Koglin}, {Madsen}, {Mao}, {Miyasaka},
  {Mori}, {Perri}, {Pivovaroff}, {Puccetti}, {Rana}, {Westergaard}, {Willis},
  {Zoglauer}, {An}, {Bachetti}, {Barri{\`e}re}, {Bellm}, {Bhalerao},
  {Brejnholt}, {Fuerst}, {Liebe}, {Markwardt}, {Nynka}, {Vogel}, {Walton},
  {Wik}, {Alexander}, {Cominsky}, {Hornschemeier}, {Hornstrup}, {Kaspi},
  {Madejski}, {Matt}, {Molendi}, {Smith}, {Tomsick}, {Ajello}, {Ballantyne},
  {Balokovi{\'c}}, {Barret}, {Bauer}, {Blandford}, {Brandt}, {Brenneman},
  {Chiang}, {Chakrabarty}, {Chenevez}, {Comastri}, {Dufour}, {Elvis}, {Fabian},
  {Farrah}, {Fryer}, {Gotthelf}, {Grindlay}, {Helfand}, {Krivonos}, {Meier},
  {Miller}, {Natalucci}, {Ogle}, {Ofek}, {Ptak}, {Reynolds}, {Rigby},
  {Tagliaferri}, {Thorsett}, {Treister}, \& {Urry}}]{NuSTAR}
{Harrison}, F.~A., {Craig}, W.~W., {Christensen}, F.~E., {et~al.} 2013, \apj,
  770, 103, \dodoi{10.1088/0004-637X/770/2/103}

\bibitem[{{Hester}(2008)}]{h2008}
{Hester}, J.~J. 2008, \araa, 46, 127,
  \dodoi{10.1146/annurev.astro.45.051806.110608}

\bibitem[{{Jansen} {et~al.}(2001){Jansen}, {Lumb}, {Altieri}, {Clavel}, {Ehle},
  {Erd}, {Gabriel}, {Guainazzi}, {Gondoin}, {Much}, {Munoz}, {Santos},
  {Schartel}, {Texier}, \& {Vacanti}}]{jla01}
{Jansen}, F., {Lumb}, D., {Altieri}, B., {et~al.} 2001, \aap, 365, L1.
\newblock
  \url{http://cdsads.u-strasbg.fr/cgi-bin/nph-bib_query?bibcode=2001A%26A...365L...1J&db_key=AST}

\bibitem[{{Kaastra} \& {Bleeker}(2016)}]{kb16}
{Kaastra}, J.~S., \& {Bleeker}, J.~A.~M. 2016, \aap, 587, A151,
  \dodoi{10.1051/0004-6361/201527395}

\bibitem[{{Kargaltsev} \& {Pavlov}(2008)}]{K&P2008}
{Kargaltsev}, O., \& {Pavlov}, G.~G. 2008, in American Institute of Physics
  Conference Series, Vol. 983, 40 Years of Pulsars: Millisecond Pulsars,
  Magnetars and More, ed. C.~{Bassa}, Z.~{Wang}, A.~{Cumming}, \& V.~M.
  {Kaspi}, 171--185, \dodoi{10.1063/1.2900138}

\bibitem[{{Madsen} {et~al.}(2015){Madsen}, {Harrison}, {Markwardt}, {An},
  {Grefenstette}, {Bachetti}, {Miyasaka}, {Kitaguchi}, {Bhalerao}, {Boggs},
  {Christensen}, {Craig}, {Forster}, {Fuerst}, {Hailey}, {Perri}, {Puccetti},
  {Rana}, {Stern}, {Walton}, {J{\o}rgen Westergaard}, \& {Zhang}}]{mhm15}
{Madsen}, K.~K., {Harrison}, F.~A., {Markwardt}, C.~B., {et~al.} 2015, \apjs,
  220, 8, \dodoi{10.1088/0067-0049/220/1/8}

\bibitem[{{Maggi} {et~al.}(2012){Maggi}, {Haberl}, {Sturm}, \& {Dewey}}]{mhs12}
{Maggi}, P., {Haberl}, F., {Sturm}, R., \& {Dewey}, D. 2012, \aap, 548, L3,
  \dodoi{10.1051/0004-6361/201220595}

\bibitem[{{Maitra} {et~al.}(2021){Maitra}, {Haberl}, {Sasaki}, {Maggi},
  {Dennerl}, \& {Freyberg}}]{mhs21}
{Maitra}, C., {Haberl}, F., {Sasaki}, M., {et~al.} 2021, arXiv e-prints,
  arXiv:2106.14532.
\newblock \doarXiv{2106.14532}

\bibitem[{{McCray}(1993)}]{mcr93}
{McCray}, R. 1993, \araa, 31, 175, \dodoi{10.1146/annurev.aa.31.090193.001135}

\bibitem[{{McCray} \& {Fransson}(2016)}]{mcf16}
{McCray}, R., \& {Fransson}, C. 2016, \araa, 54, 19,
  \dodoi{10.1146/annurev-astro-082615-105405}

\bibitem[{{Miceli} {et~al.}(2019){Miceli}, {Orlando}, {Burrows}, {Frank},
  {Argiroffi}, {Reale}, {Peres}, {Petruk}, \& {Bocchino}}]{mob19}
{Miceli}, M., {Orlando}, S., {Burrows}, D.~N., {et~al.} 2019, Nature Astronomy,
  3, 236, \dodoi{10.1038/s41550-018-0677-8}

\bibitem[{{Mignone} {et~al.}(2012){Mignone}, {Zanni}, {Tzeferacos}, {van
  Straalen}, {Colella}, \& {Bodo}}]{2012ApJS..198....7M}
{Mignone}, A., {Zanni}, C., {Tzeferacos}, P., {et~al.} 2012, \apjs, 198, 7,
  \dodoi{10.1088/0067-0049/198/1/7}

\bibitem[{{Ono} {et~al.}(2020){Ono}, {Nagataki}, {Ferrand}, {Takahashi},
  {Umeda}, {Yoshida}, {Orlando}, \& {Miceli}}]{onf20}
{Ono}, M., {Nagataki}, S., {Ferrand}, G., {et~al.} 2020, \apj, 888, 111,
  \dodoi{10.3847/1538-4357/ab5dba}

\bibitem[{{Orlando} {et~al.}(2015){Orlando}, {Miceli}, {Pumo}, \&
  {Bocchino}}]{omp15}
{Orlando}, S., {Miceli}, M., {Pumo}, M.~L., \& {Bocchino}, F. 2015, \apj, 810,
  168, \dodoi{10.1088/0004-637X/810/2/168}

\bibitem[{{Orlando} {et~al.}(2020){Orlando}, {Ono}, {Nagataki}, {Miceli},
  {Umeda}, {Ferrand}, {Bocchino}, {Petruk}, {Peres}, {Takahashi}, \&
  {Yoshida}}]{oon20}
{Orlando}, S., {Ono}, M., {Nagataki}, S., {et~al.} 2020, \aap, 636, A22,
  \dodoi{10.1051/0004-6361/201936718}

\bibitem[{{Page} {et~al.}(2020){Page}, {Beznogov}, {Garibay}, {Lattimer},
  {Prakash}, \& {Janka}}]{pbg20}
{Page}, D., {Beznogov}, M.~V., {Garibay}, I., {et~al.} 2020, arXiv e-prints,
  arXiv:2004.06078.
\newblock \doarXiv{2004.06078}

\bibitem[{{Panagia}(1999)}]{pan99}
{Panagia}, N. 1999, in New Views of the Magellanic Clouds, ed. Y.~H. {Chu},
  N.~{Suntzeff}, J.~{Hesser}, \& D.~{Bohlender}, Vol. 190, 549

\bibitem[{{Park} {et~al.}(2006){Park}, {Zhekov}, {Burrows}, {Garmire},
  {Racusin}, \& {McCray}}]{pzb06}
{Park}, S., {Zhekov}, S.~A., {Burrows}, D.~N., {et~al.} 2006, \apj, 646, 1001,
  \dodoi{10.1086/505023}

\bibitem[{{Ravi} {et~al.}(2021){Ravi}, {Park}, {Zhekov}, {Miceli}, {Orlando},
  {Frank}, \& {Burrows}}]{rpz21}
{Ravi}, A.~P., {Park}, S., {Zhekov}, S.~A., {et~al.} 2021, arXiv e-prints,
  arXiv:2109.02881.
\newblock \doarXiv{2109.02881}

\bibitem[{{Reynolds} {et~al.}(2017){Reynolds}, {Pavlov}, {Kargaltsev},
  {Klingler}, {Renaud}, \& {Mereghetti}}]{Reynolds:2017}
{Reynolds}, S.~P., {Pavlov}, G.~G., {Kargaltsev}, O., {et~al.} 2017, \ssr, 207,
  175, \dodoi{10.1007/s11214-017-0356-6}

\bibitem[{{Russell} \& {Dopita}(1992)}]{rd92}
{Russell}, S.~C., \& {Dopita}, M.~A. 1992, \apj, 384, 508,
  \dodoi{10.1086/170893}

\bibitem[{{Slane}(2017)}]{Slane:2017}
{Slane}, P. 2017, in Handbook of Supernovae, ed. A.~W. {Alsabti} \&
  P.~{Murdin}, 2159, \dodoi{10.1007/978-3-319-21846-5\_95}

\bibitem[{{Sugerman} {et~al.}(2005){Sugerman}, {Crotts}, {Kunkel}, {Heathcote},
  \& {Lawrence}}]{sck05}
{Sugerman}, B. E.~K., {Crotts}, A. P.~S., {Kunkel}, W.~E., {Heathcote}, S.~R.,
  \& {Lawrence}, S.~S. 2005, \apjs, 159, 60, \dodoi{10.1086/430408}

\bibitem[{{Sun} {et~al.}(2021){Sun}, {Vink}, {Chen}, {Zhou}, {Prokhorov},
  {Puhlhofer}, \& {Malyshev}}]{svc21}
{Sun}, L., {Vink}, J., {Chen}, Y., {et~al.} 2021, arXiv e-prints,
  arXiv:2103.03844.
\newblock \doarXiv{2103.03844}

\bibitem[{{Torres} {et~al.}(2014){Torres}, {Cillis}, {Mart{\'{\i}}n}, \& {de
  O{\~n}a Wilhelmi}}]{Torres:2014}
{Torres}, D.~F., {Cillis}, A., {Mart{\'{\i}}n}, J., \& {de O{\~n}a Wilhelmi},
  E. 2014, ArXiv e-prints.
\newblock \doarXiv{1402.5485}

\bibitem[{{Urushibata} {et~al.}(2018){Urushibata}, {Takahashi}, {Umeda}, \&
  {Yoshida}}]{2018MNRAS.473L.101U}
{Urushibata}, T., {Takahashi}, K., {Umeda}, H., \& {Yoshida}, T. 2018, \mnras,
  473, L101, \dodoi{10.1093/mnrasl/slx166}

\bibitem[{{Vissani}(2015)}]{vis15}
{Vissani}, F. 2015, Journal of Physics G Nuclear Physics, 42, 013001,
  \dodoi{10.1088/0954-3899/42/1/013001}

\bibitem[{{West} {et~al.}(1987){West}, {Lauberts}, {Jorgensen}, \&
  {Schuster}}]{wls87}
{West}, R.~M., {Lauberts}, A., {Jorgensen}, H.~E., \& {Schuster}, H.~E. 1987,
  \aap, 177, L1

\bibitem[{{Wilms} {et~al.}(2000){Wilms}, {Allen}, \& {McCray}}]{wam00}
{Wilms}, J., {Allen}, A., \& {McCray}, R. 2000, \apj, 542, 914,
  \dodoi{10.1086/317016}

\bibitem[{{Zhekov} {et~al.}(2009){Zhekov}, {McCray}, {Dewey}, {Canizares},
  {Borkowski}, {Burrows}, \& {Park}}]{zmd09}
{Zhekov}, S.~A., {McCray}, R., {Dewey}, D., {et~al.} 2009, \apj, 692, 1190,
  \dodoi{10.1088/0004-637X/692/2/1190}

\end{thebibliography}
\bibliographystyle{aasjournal}

\newpage
\clearpage
\appendix

\section{PWN emission in the soft X-rays} \label{app:absorbed_spec}

The thermal emission of \sna\ has been increasing with time in the last two decades. Therefore, the non-thermal emission, expected to be basically constant, should be easier to detect in the earlier epochs, when the contamination by the thermal emission was smaller. However, the absorption of the cold ejecta is significantly higher in the early stage of the evolution of \sna\ and it is not obvious to figure out which of the two effects (either reduced contamination by thermal emission or increased absorption) is the most relevant for the global emission. What we know is that no significant X-ray non-thermal emission has ever been detected in the band between 0.5 and 8 keV. Therefore, to test the scenario presented in this paper, we checked whether the PWN we detected is compatible with the \chandra\ observations performed before the launch of \nustar. In particular, we considered data collected by \chandra/ACIS-S detector in 2003, 2014 and 2020 (see Table~\ref{tab:obs}). We extracted spectra from the inner part of \sna, black circle of 0.''3 radius, focusing on the faintest region where the NS is expected to lie. Since the inner area of \sna\ is very faint, we combined the 2014 and 2020 spectra through the task \texttt{addascaspec} to decrease the width of the error bars.

\begin{figure*}[!ht]
    \centering
    \begin{minipage}{0.3\textwidth}
    \includegraphics[width=\textwidth]{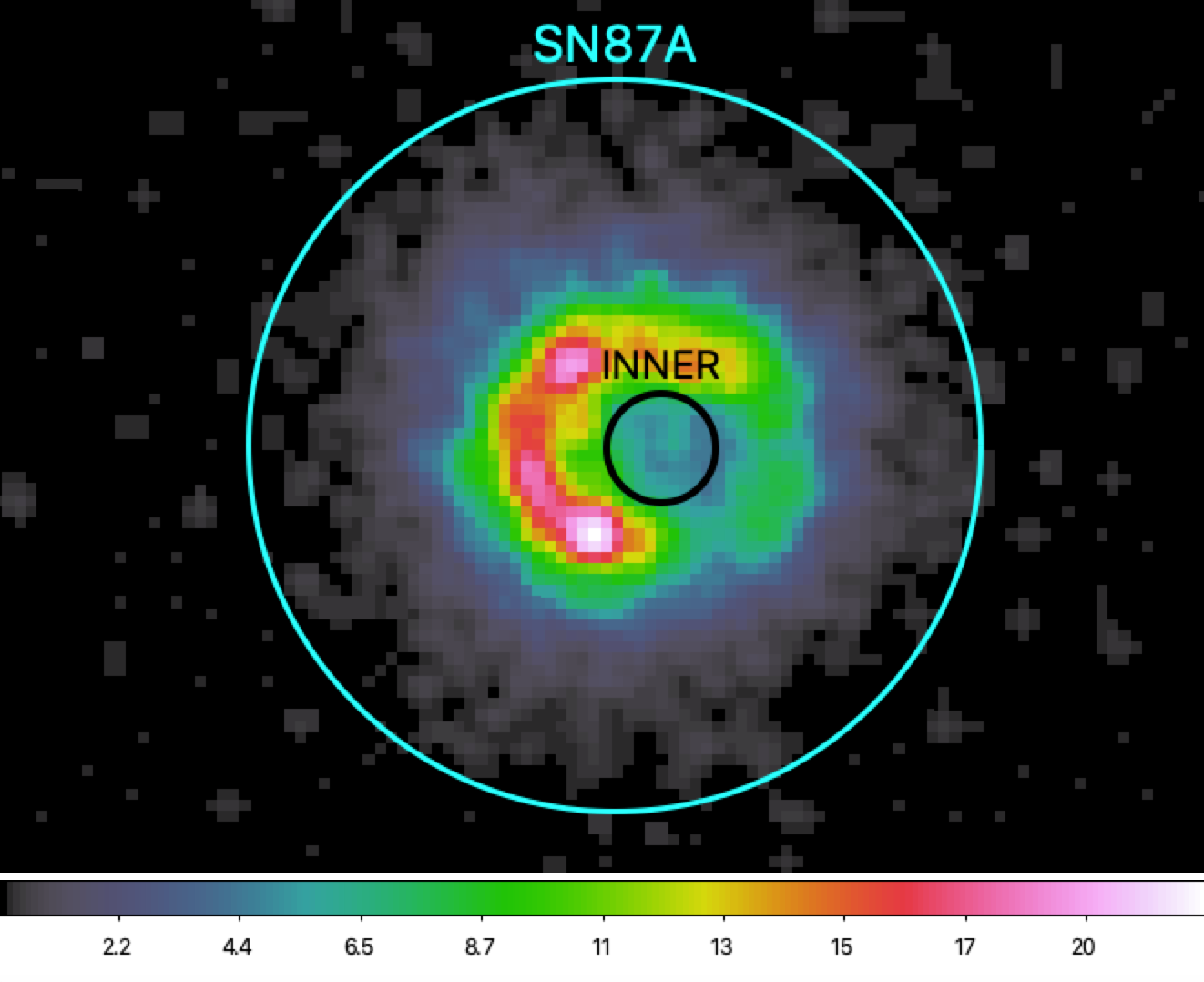}
    \end{minipage}
    \hfill
    \begin{minipage}{0.3\textwidth}
    \includegraphics[width=\textwidth]{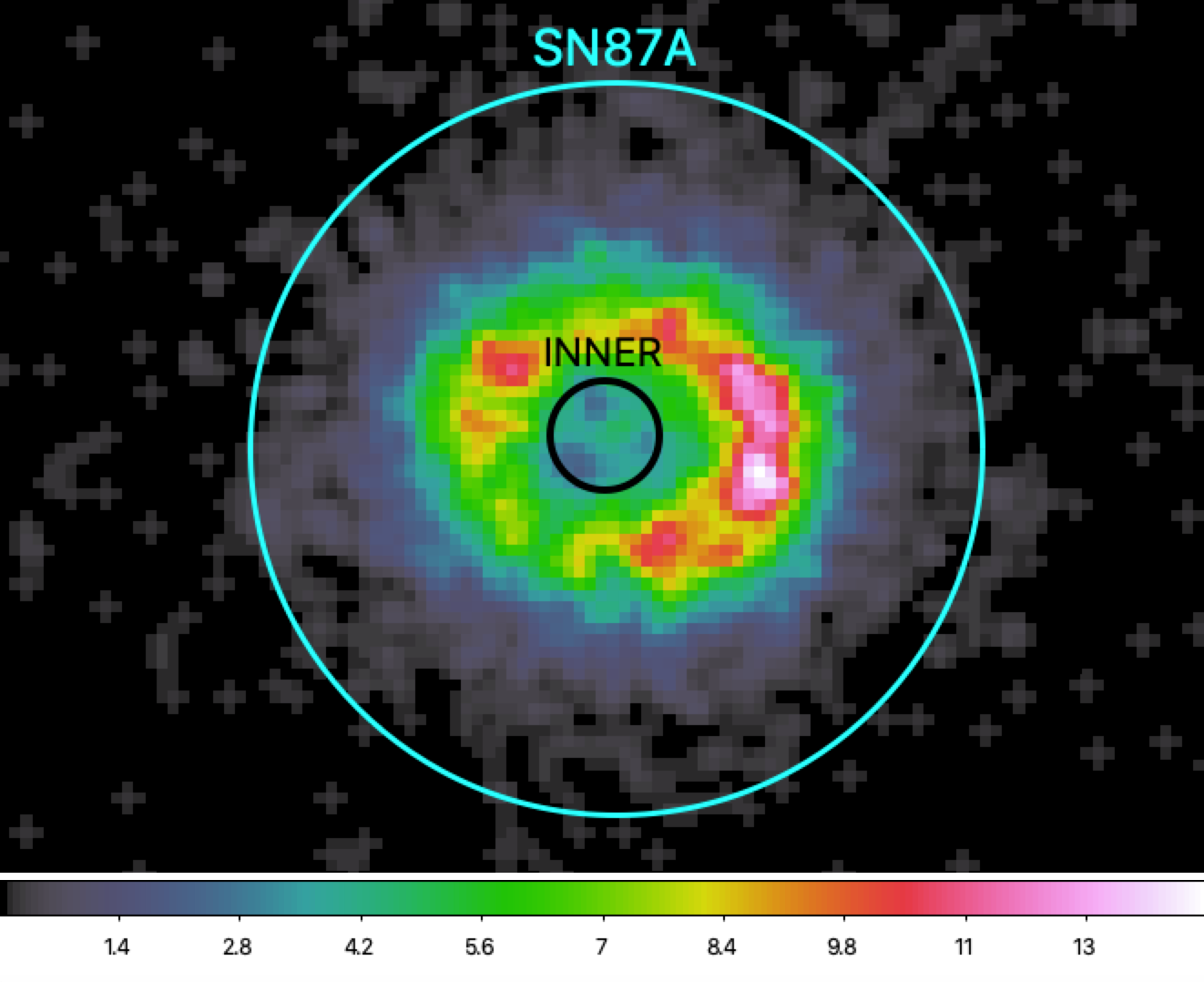}
    \end{minipage}
    \hfill
    \begin{minipage}{0.3\textwidth}
    \includegraphics[width=\textwidth]{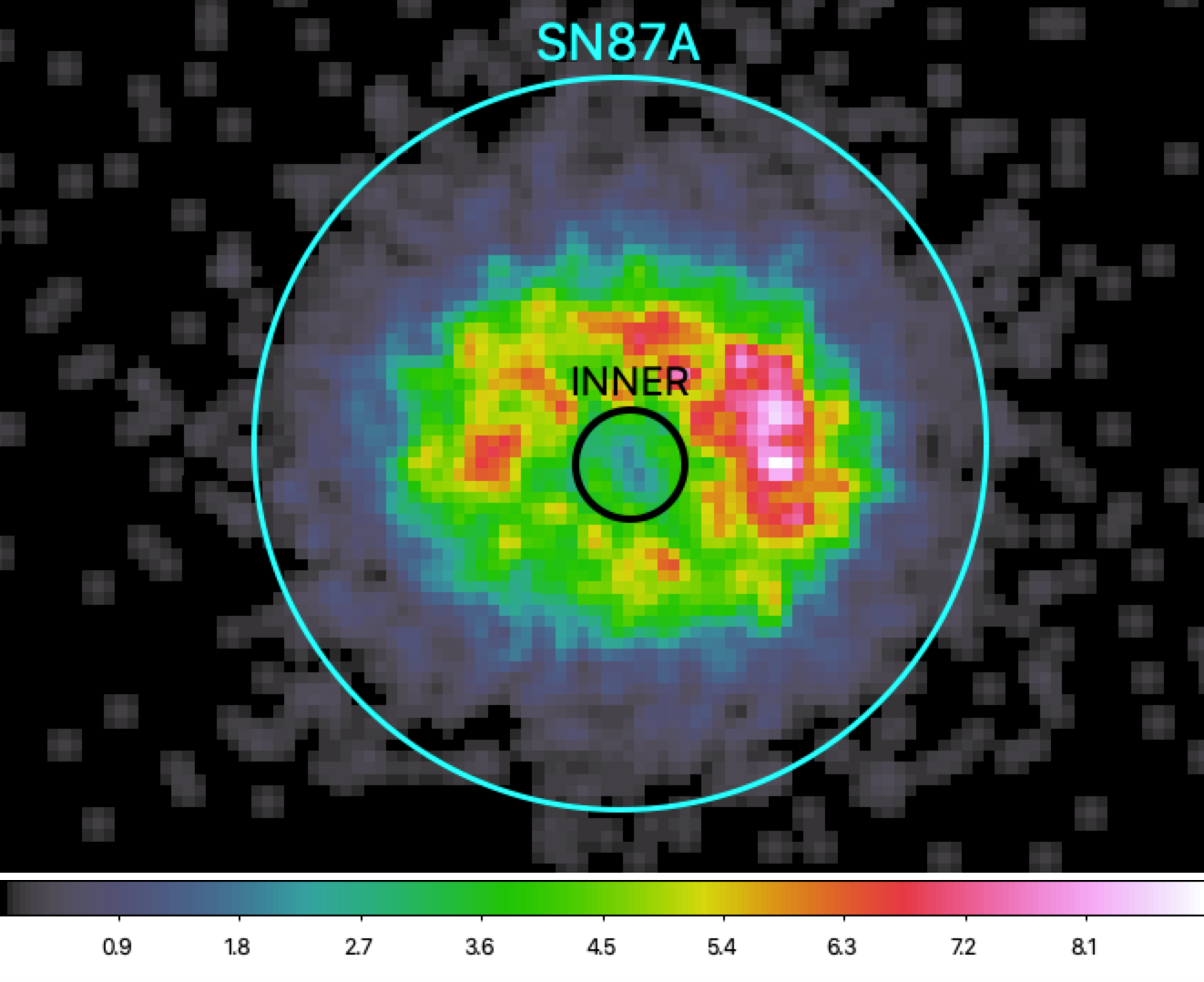}
    \end{minipage}
    \begin{minipage}{0.32\textwidth}
    \includegraphics[angle=270,width=\textwidth]{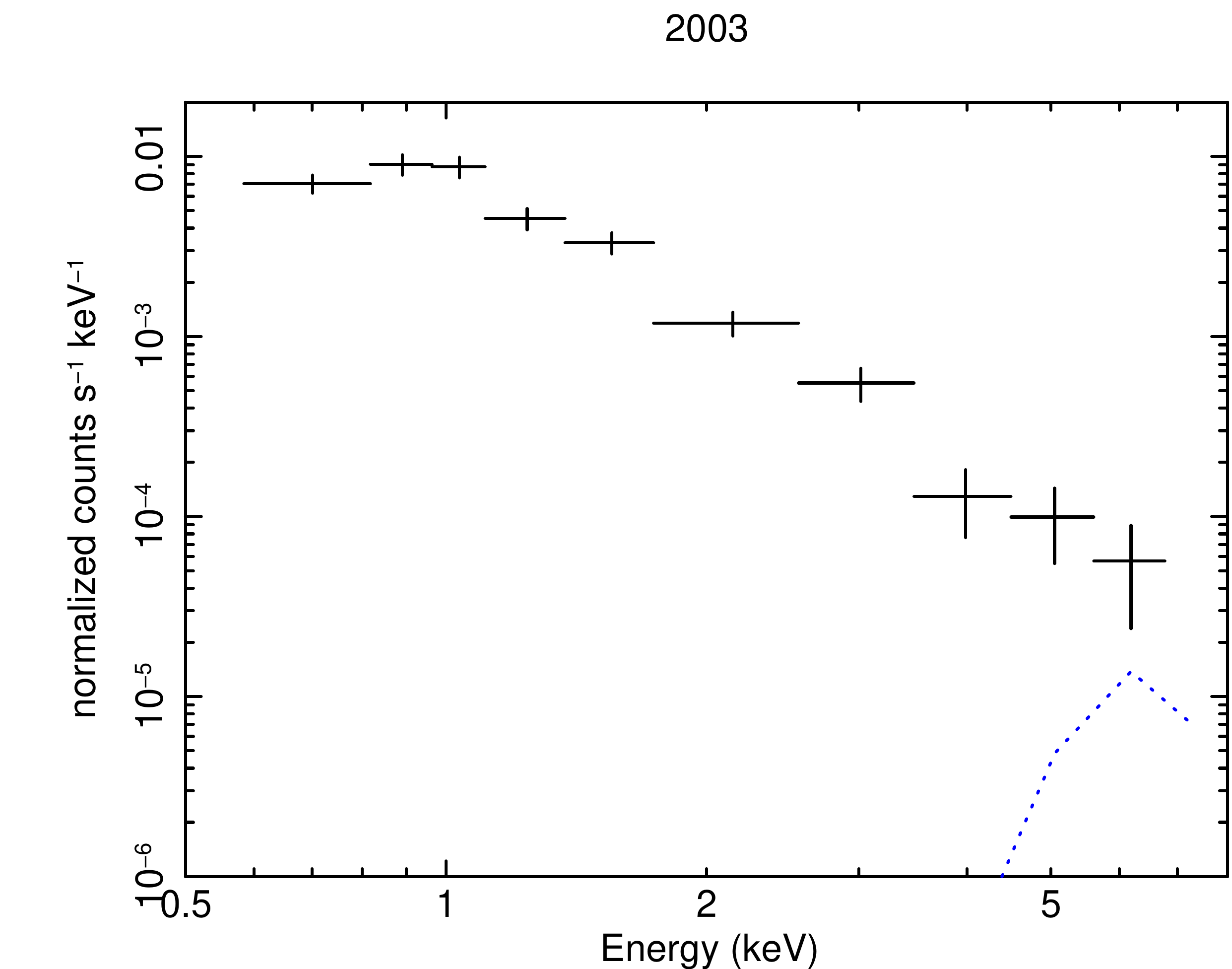}
    \end{minipage}
    \hfill
    \begin{minipage}{0.32\textwidth}
    \includegraphics[angle=270,width=\textwidth]{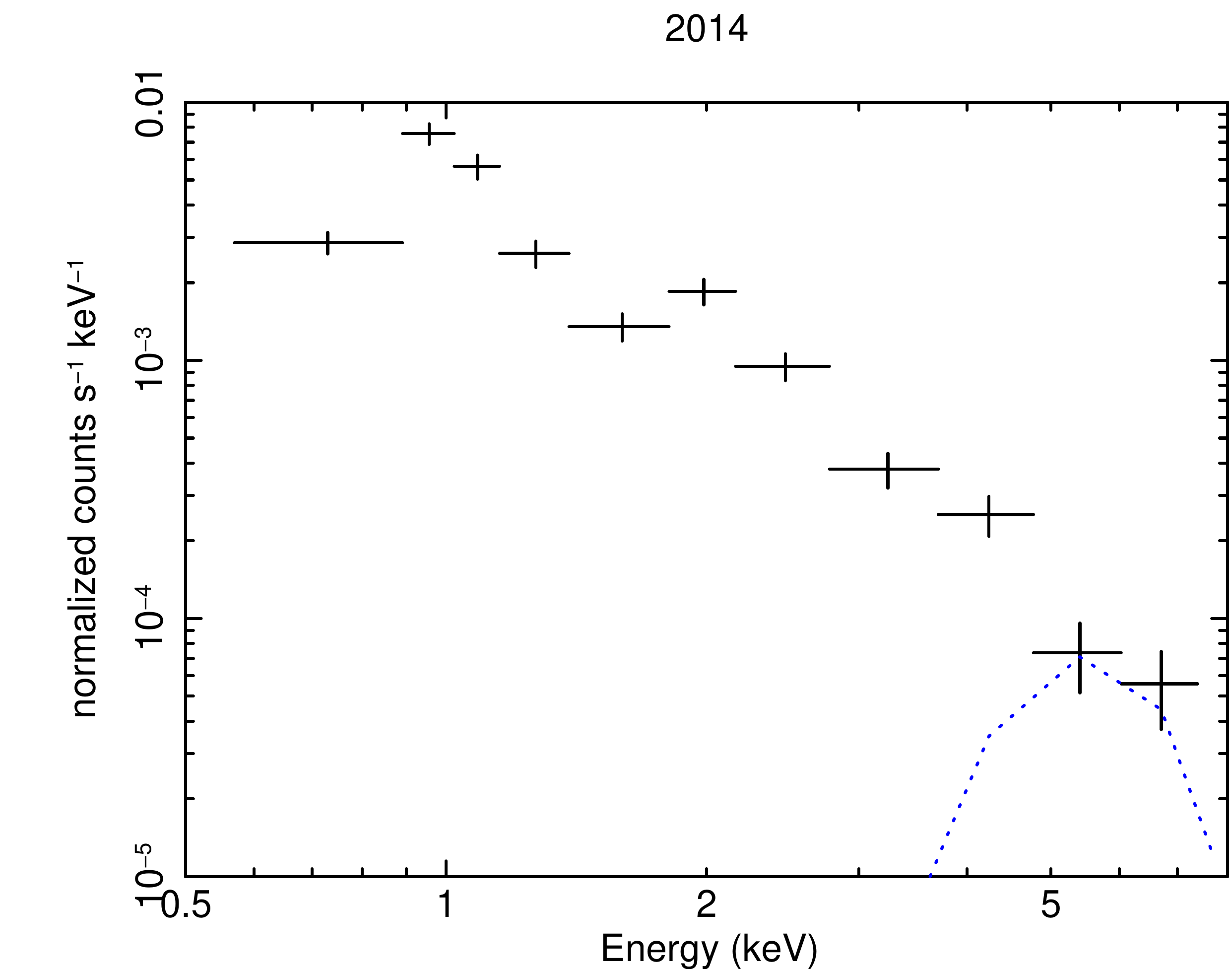}
    \end{minipage}
    \hfill
    \begin{minipage}{0.32\textwidth}
    \includegraphics[angle=270,width=\textwidth]{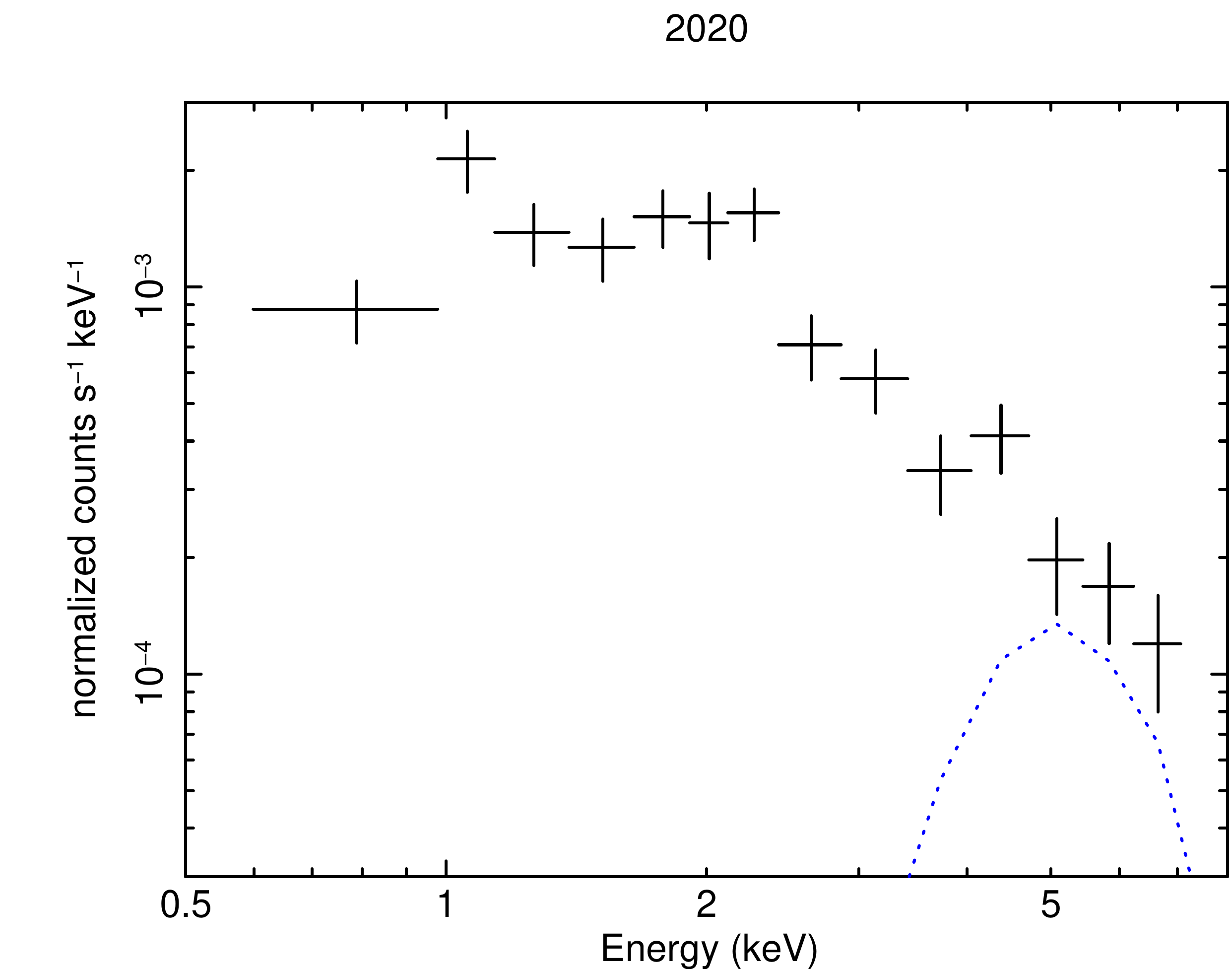}
    \end{minipage}
    \caption{\emph{First row.} \chandra/ACIS-S count images of \sna\ in the 0.1-8 keV band in 2003 (left), 2014 (center) and 2020 (right). The images are smoothed with a gaussian having a sigma of 1 pixel. The black and cyan circle mark the the inner faint region of \sna, where the putative compact object is most likely lying, and the whole remnant. \emph{Second row.} Combined spectra of \sna\ extracted from the \chandra\ observations performed in 2003 (left), 2014 (center) and 2020 (right). The black points stand for the inner region and the blue curve shows the emission of the absorbed PL.}
    \label{fig:abs_pwn_past}
\end{figure*}

 The spectra are then optimally binned and compared with the emission of the PWN. We assumed photon index $\Gamma=2.8$ and norm$=7.1 \times 10^{-4}$ ph/s/keV/cm$^{2}$, namely the PL parameters reported in Table \ref{tab:fit_multiepoch}. The PL component is coupled to the proper \texttt{vphabs} component for the considered year, estimated from model B18.3. Since the inner faint part of \sna\ is smaller than the \chandra/ACIS-S pixel size (0.''3 vs 0.''492), the observed flux must be corrected to take into account the PSF effects, particularly the encircled energy. Therefore, we included a multiplicative constant in the spectra of the PWN when it is compared to the spectrum of the inner region. For a radius of 0.''3 and assuming the peak energy around 6 keV, this constant is $\sim 0.4$ (https://cxc.harvard.edu/proposer/POG/pdf/MPOG.pdf). The resulting spectra for all the years are shown in the second row of Fig. \ref{fig:abs_pwn_past}. 

As it is clear from the plots in Fig. \ref{fig:abs_pwn_past}, in every year, the non-thermal emission is at least a factor $\sim 10$ dimmer than the thermal one up to 5 keV. At energies $\sim$5 keV, the absorbed PL reaches its peak but, in any case,  it is  below the observed spectrum.  

Moreover, the emission observed in the inner area is due to the contamination of the shocked ring material and the potential signal from the PWN would be mixed up with this contamination. At energies $\gtrsim$ 5 keV, the flux of the observed spectrum dramatically decreases and becomes compatible with the background. For all these reasons, the PWN is currently undetectable in the soft (0.5-8 keV) X-ray band.

\end{document}